\documentclass[preprint,superscriptaddress,nofootinbib]{revtex4-1}
\usepackage{graphicx}
\usepackage{dcolumn}
\usepackage{bm}
\usepackage{amsmath}
\usepackage{amsfonts} 
\usepackage{latexsym}
\usepackage{bbm}
\usepackage{color}
\usepackage{amssymb}
\usepackage{amsthm}
\usepackage{epsf}
\usepackage{slashed}
\usepackage{epsfig}
\usepackage{caption3}
\usepackage{appendix}
\usepackage{cases}
\usepackage{subfigure}
\usepackage{multirow}
\usepackage{amsthm,amsmath,amssymb}
\usepackage{mathrsfs}
\usepackage{tcolorbox}
\usepackage{colortbl}

\makeatletter

\newcommand{\Rmnum}[1]{\expandafter\@slowromancap\romannumeral #1@}
\makeatother

\begin{document}
	
\title{Hydrodynamic effects on the filtered dark matter produced by a first-order phase transition}
	
\author{Siyu Jiang}

\author{Fa Peng Huang}%
\email{Corresponding author. huangfp8@sysu.edu.cn}

\affiliation{MOE Key Laboratory of TianQin Mission, TianQin Research Center for
	Gravitational Physics \& School of Physics and Astronomy, Frontiers
	Science Center for TianQin, Gravitational Wave Research Center of CNSA, 
	Sun Yat-sen University (Zhuhai Campus), Zhuhai 519082, China}

\author{Chong Sheng Li}
\affiliation{School of Physics and State Key Laboratory of Nuclear Physics and Technology, Peking University, Beijing 100871, China}
\affiliation{Center for High Energy Physics, Peking University, Beijing 100871, China}
\bigskip
	
	
\begin{abstract}
Motivated by current status of dark matter (DM) search, a new type of DM production mechanism is proposed based on the dynamical process of a strong first-order phase transition in the early Universe, namely, the filtered DM mechanism. We study the hydrodynamic effects on the DM relic density.  By detailed calculations, we demonstrate that the hydrodynamic modes with the corresponding hydrodynamic heating effects play essential roles in determining the DM relic density. The corresponding phase transition gravitational wave could help to probe this new mechanism.
\end{abstract}

\maketitle

\section{Introduction}
Exploring the microscopic nature of dark matter (DM)~\cite{Bertone:2016nfn} is an important goal in the interplay of particle physics and cosmology. Besides the well-studied freeze-out, freeze-in and misalignment DM formation mechanisms,
recently a new type of mechanism is proposed that the DM is produced by the dynamical process during 
a strong first-order phase transition (SFOPT)~\cite{Baker:2019ndr,Chway:2019kft, Krylov:2013qe, Huang:2017kzu}. This mechanism could naturally evade the unitarity
constraints to form the heavy DM~\cite{Griest:1989wd,Smirnov:2019ngs} and can be detected by phase transition gravitational wave (GW).
Especially, the intriguing filtered DM mechanism proposed in Refs.~\cite{Baker:2019ndr,Chway:2019kft} assumes that the DM candidate particles acquire masses when they enter into the bubbles and too heavy DM particles are energetically unfavorable to enter into the broken phase. Thus only an extremely small fraction of DM particles finally enter into the bubbles and then contribute to the observed DM relic density. This mechanism applies to not only the DM model but also baryogenesis~\cite{Baldes:2021vyz,Marfatia:2020bcs,Ahmadvand:2021vxs,Chao:2020adk,Azatov:2021ifm,Baker:2021zsf,Azatov:2021irb,Huang:2022vkf}.
In the original work of filtered DM, the hydrodynamic
effects in the vicinity of the bubble wall are not considered. For example, 
the temperature in front of and behind the bubble walls is chosen as the same nucleation temperature. However, the DM density depends on the temperature and velocity distribution. 
Actually, to realize the filtered DM, usually it should be a SFOPT with a large phase transition strength parameter $\alpha$, which leads to obvious temperature and velocity differences in front of and behind the bubble wall. These differences originate from the hydrodynamic effects and can significantly enhance or suppress the DM relic density.
We study these
hydrodynamic effects with dedicated calculations and find that the DM relic density is sensitive to
the hydrodynamic modes and the corresponding heating effects. 

In Sec.~\ref{anaestimation},  we simply present the analytic estimation for the filtered DM mechanism and clearly show the reason to consider the hydrodynamic effects. The phase transition dynamics including the bubble wall velocity are discussed in Sec.~\ref{PTdynamics}. In Sec.~\ref{hydro}, we show how to calculate hydrodynamic profiles of temperature and velocity for zero-width bubble wall. We present the calculations both for spherical and planar bubble wall. The hydrodynamic effects on the analytic DM relic density are also given and we emphasize the accuracy of simple low-velocity approximation in the deflagration mode. In Sec.~\ref{Tvprofile} we show the temperature and velocity profiles across the bubble wall, which has a nonzero width. In Sec.~\ref{boltzmann} we solve the Boltzmann equation numerically for bubble wall with finite width to get more accurate results. The results of hydrodynamic effects on DM relic density for four benchmark points are also given. The phase transition GW signals of the filtered DM are given in Sec.~\ref{gw}. Concise conclusions are given in Sec.~\ref{conclusion}.

\section{analytic estimation}\label{anaestimation}
To clearly see that the hydrodynamic effects  play important roles in the final 
DM relic density, we first demonstrate the simple analytic estimation of the DM density.
The detailed numerical calculation of the DM density by using the Boltzmann
equations to include the hydrodynamic effects is also performed in Sec.~\ref{boltzmann}.

\begin{figure}[htbp]
	\begin{center}
		\includegraphics[width=0.7\linewidth]{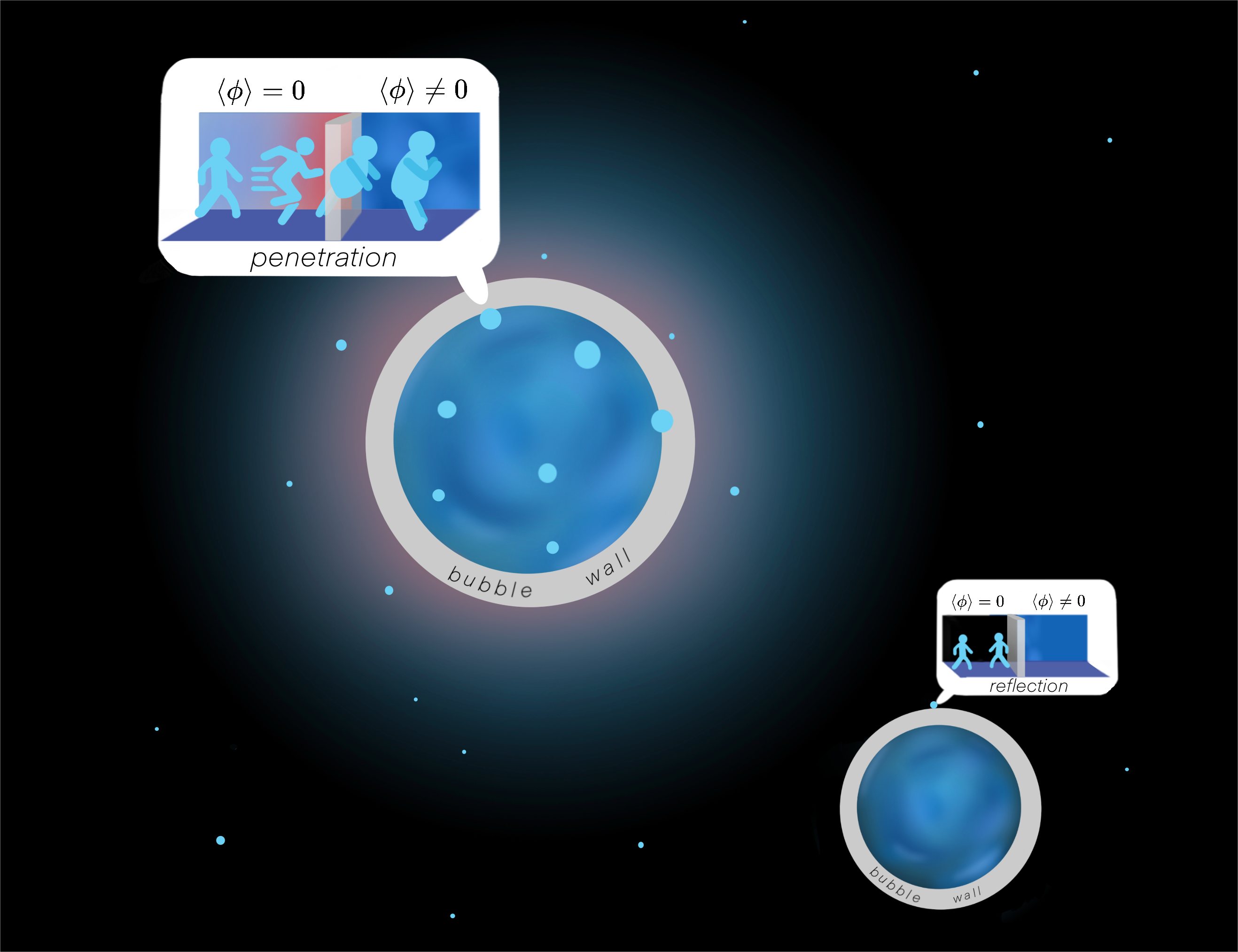}
		\caption{Schematic process for the filtered DM with hydrodynamic effects for the deflagration mode. In upper left we show that due to the heating effects the DM has more abilities to pass through the bubble wall. In lower right the DM more likely reflects from the bubble wall.}
		\label{fig_cartoon}
	\end{center}
\end{figure}
In Fig.~\ref{fig_cartoon},  we show the basic process of the filtered DM with hydrodynamic effects. $\phi$ is the phase transition field during a SFOPT, and bubbles nucleate in the thermal plasma of the early Universe. In the upper left we show the deflagration mode, which is due to the hydrodynamic effects of the region in front of the bubble wall heating (red color) and then the DM particles increase the probability to pass through the bubble wall than one without considering hydrodynamic effects. In the lower right we show the case without heating or the case for the detonation of DM, which more likely reflects from the bubble wall.
The bubble walls formed during a SFOPT can provide new approaches to produce heavy DM through the filtering effects.
Due to the energy conservation, only a Boltzmann-suppressed fraction of the wouldbe DM particles with very large momenta can pass through the bubbles walls and then contribute to the 
DM relic density when the particles hit the wall.
It is straightforward to discuss the filtering effects in the bubble wall frame with the following condition~\cite{Baker:2019ndr,Chway:2019kft}:
\begin{eqnarray}
p_z^w>\sqrt{\Delta m^2}\,\,,
\end{eqnarray}
where $p_z^w$ is the particle $z$-direction momentum in the bubble wall frame, $\Delta m^2=(m_\chi^{\mathrm{in}})^2-m_0^2$ with $m_\chi^{\mathrm{in}}$ the mass of DM particle $\chi$ deep inside the bubble and $m_0$ the mass in the false vacuum~\cite{Chao:2020adk}.

Assuming the DM particle $\chi$ is in thermal equilibrium outside the bubble, such that in the bubble wall frame its distribution function follows the Bose-Einstein or Fermi-Dirac distribution, then
\begin{eqnarray}\label{feq}
f_{\chi}^{\rm eq}=\frac{1}{e^{\tilde \gamma_{\rm pl}\left(\sqrt{({{p}^w})^2+m_0^2}-\tilde v_{\rm pl} {p}_z^w\right) / T}\mp 1}\,\,,
\end{eqnarray}
where $\tilde v_{\rm pl}$ is the relative velocity of fluid bulk motion with respect to the wall and $\tilde \gamma_{\rm pl}=1/\sqrt{1-\tilde v_{\mathrm{pl}}^2}$ is its Lorentz factor. $p^w\equiv |\mathbf{p}^w|$ is the magnitude of the three-momentum of DM and $T$ is the temperature of DM. The particle flux coming from the false vacuum per unit area and unit time can be written as~\cite{Baker:2019ndr,Chway:2019kft}
\begin{eqnarray}\label{flux}
{J}_{\chi}^w=g_\chi \int \frac{d^3 {p^w}}{(2 \pi)^3} \frac{{p}_z^w}{{E}^w} f_\chi^{\rm eq} \Theta\left({p}_z^w-\sqrt{\Delta m^2}\right) \,\,,
\end{eqnarray}
where $g_{\chi}$ is the DM degrees of freedom. 
Then the DM number density inside the bubble $n_{\chi}^{\text {in }}$ in the bubble center frame can be written as~\cite{Baker:2019ndr,Chway:2019kft} 
\begin{eqnarray}\label{nin}
n_{\chi}^{\text {in }}=\frac{{J}_{\chi}^w}{  \gamma_{w}   v_{w}}\,\,,
\end{eqnarray}
where $v_w$ is the bubble wall velocity and $\gamma_w=1/\sqrt{1-v_w^2}$ is its Lorentz factor. In our work we set $m_0=0$, and then we can integrate Eq.~\eqref{flux} and get~\cite{Chway:2019kft}
\begin{eqnarray}\label{nana1}
	n_\chi^{\text {in }} \simeq \frac{g_{\chi} T_{}^3}{\gamma_w v_w}\left(\frac{\tilde \gamma_{\mathrm{pl}}\left(1- \tilde v_{\mathrm{pl}}\right) m_\chi^{\mathrm{in}} / T_{}+1}{4 \pi^2 \tilde \gamma_{\mathrm{pl}}^3\left(1- \tilde v_{\mathrm{pl}}\right)^2}\right) e^{-\frac{\tilde \gamma_{\mathrm{pl}}\left(1- \tilde v_{\mathrm{pl}}\right) m_\chi^{\mathrm{in}}}{T_{}}} \,\,,
\end{eqnarray}
where we have used Maxwell-Boltzmann approximation of DM distribution.
One can see that, as $\tilde{v}_{\mathrm{pl}} \rightarrow 1, \tilde{\gamma}_\mathrm{pl} \gg 1$, by using $\tilde \gamma_\mathrm{pl}\left(1-\tilde v_\mathrm{pl}\right)=\tilde \gamma_\mathrm{pl}-\sqrt{\tilde \gamma_\mathrm{pl}^2-1} \rightarrow \frac{1}{2 \tilde \gamma_\mathrm{pl}}$, the exponent approaches $-m_\chi^{\mathrm{in}} / (2 \tilde{\gamma}_\mathrm{pl} T)$. On the other hand, in the case of $\tilde{v}_\mathrm{pl} \rightarrow 0, \tilde{\gamma}_\mathrm{pl} \rightarrow 1$, the exponent becomes $-m_\chi^{\mathrm{in}} / T$. We show that, when $\tilde \gamma_\mathrm{pl}\gg m_\chi^{\mathrm{in}} /T$, Eq.~\eqref{nana1} approaches $g_\chi T^3 / \pi^2$, which is the equilibrium number density for Boltzmann distribution outside the bubble. It means that all particles can pass through the bubble wall, i.e., the bubble wall does not filter out DM particles in this limit.

The filtering effect is mainly represented by the exponential term of Eq.~\eqref{nana1}. When the fluid velocity in the wall frame is low, the kinetic energy of DM particles is $\mathcal{O}(T)$. When the fluid velocity is relativistic, the kinetic energy of DM particles is $\mathcal{O}(\tilde \gamma_{\mathrm{pl}}T)$. Only small fraction of particles can penetrate into the bubble when their kinetic energy is smaller than their mass deep inside the bubble $m_\chi^{\mathrm{in}}$. The DM number density inside the bubble is sensitive to the velocity, temperature, and mass of DM since they appear in the exponential term.

The DM abundance today can be calculated by dividing $n_{\chi}^{\text {in}}$ by the entropy density $s=\left(2 \pi^2 / 45\right) g_{\star}(T') T'^3$ inside the bubble, where $g_{\star}(T')$ is the effective number of relativistic degrees of freedom associated with entropy and $T'$ is the temperature behind the bubble wall. After normalizing to the critical density $\rho_c=3H_0^2 M_{\mathrm{pl}}^2$ we have~\cite{Baker:2019ndr,Marfatia:2020bcs} :
\begin{eqnarray}\label{oh20}
	\Omega_{\mathrm{DM}} h^2 =\frac{m_{\chi}^{\mathrm{in}} (n_\chi^{\rm in}+n_{\bar\chi}^{\rm in})}{\rho_c / h^2} \frac{ g_{\star 0} T_0^3}{ g_{\star }\left(T'\right) T'^3}\simeq 6.29 \times 10^8 \frac{m_{\chi}^{\mathrm{in}}}{\mathrm{GeV}} \frac{(n_\chi^{\rm in}+n_{\bar\chi}^{\rm in})}{g_{\star}\left(T'\right) T'^3}\,\,.
\end{eqnarray}
where $T_0\simeq 0.235~\mathrm{meV}$ is the temperature today and $g_{\star 0}=3.9$ is the effective degree of freedom at present. 
$h = H_0/(100~\mathrm{km/sec/Mpc}) = 0.678$~\cite{Planck:2018vyg} is the dimensionless parameter in the observed Hubble constant $H_0$. 

\begin{figure}[htbp]
	\centering
	\subfigure{
		\begin{minipage}[t]{0.5\linewidth}
			\centering
			\includegraphics[scale=0.52]{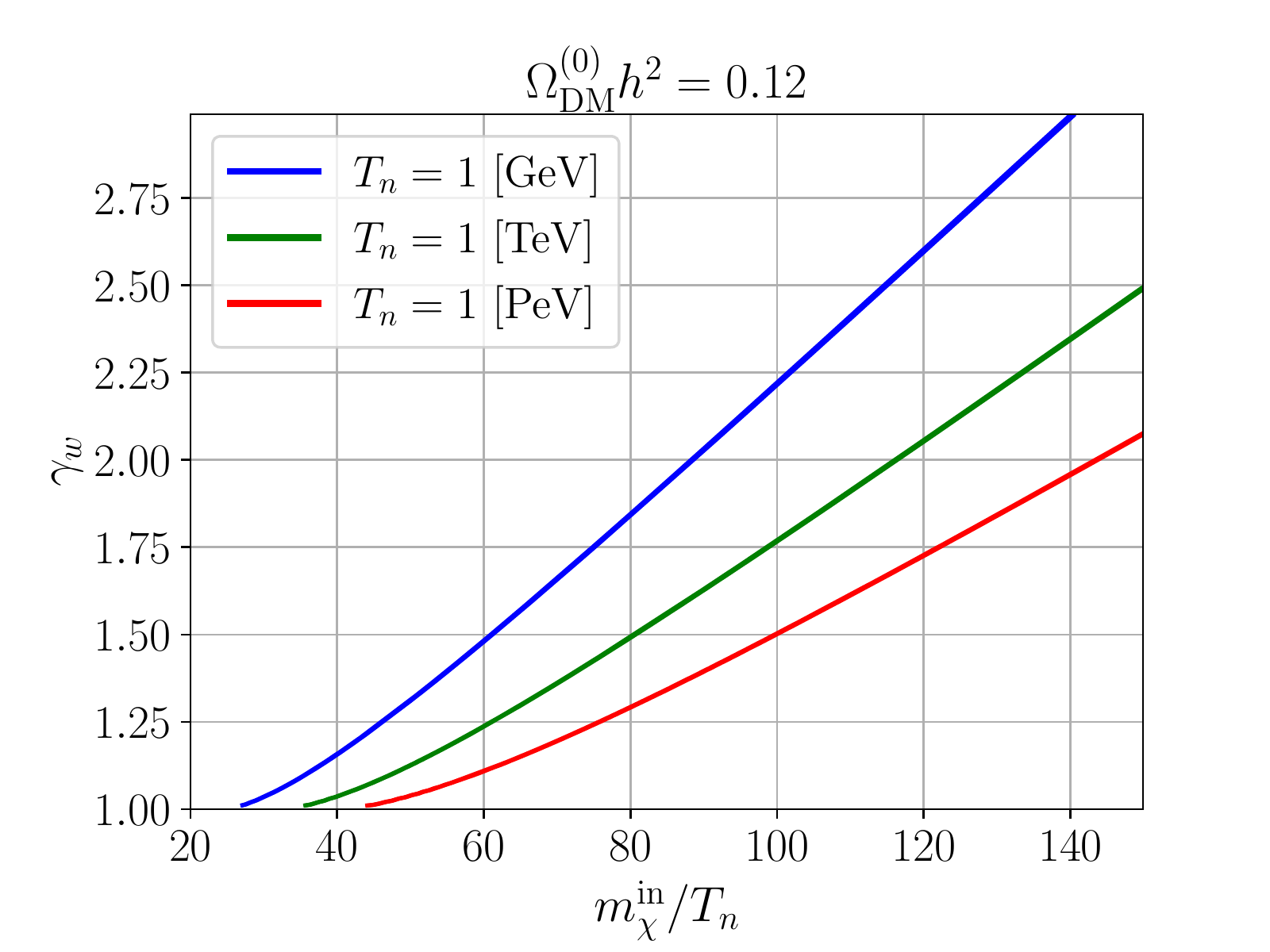}
	\end{minipage}}%
	\subfigure{
		\begin{minipage}[t]{0.5\linewidth}
			\centering
			\includegraphics[scale=0.52]{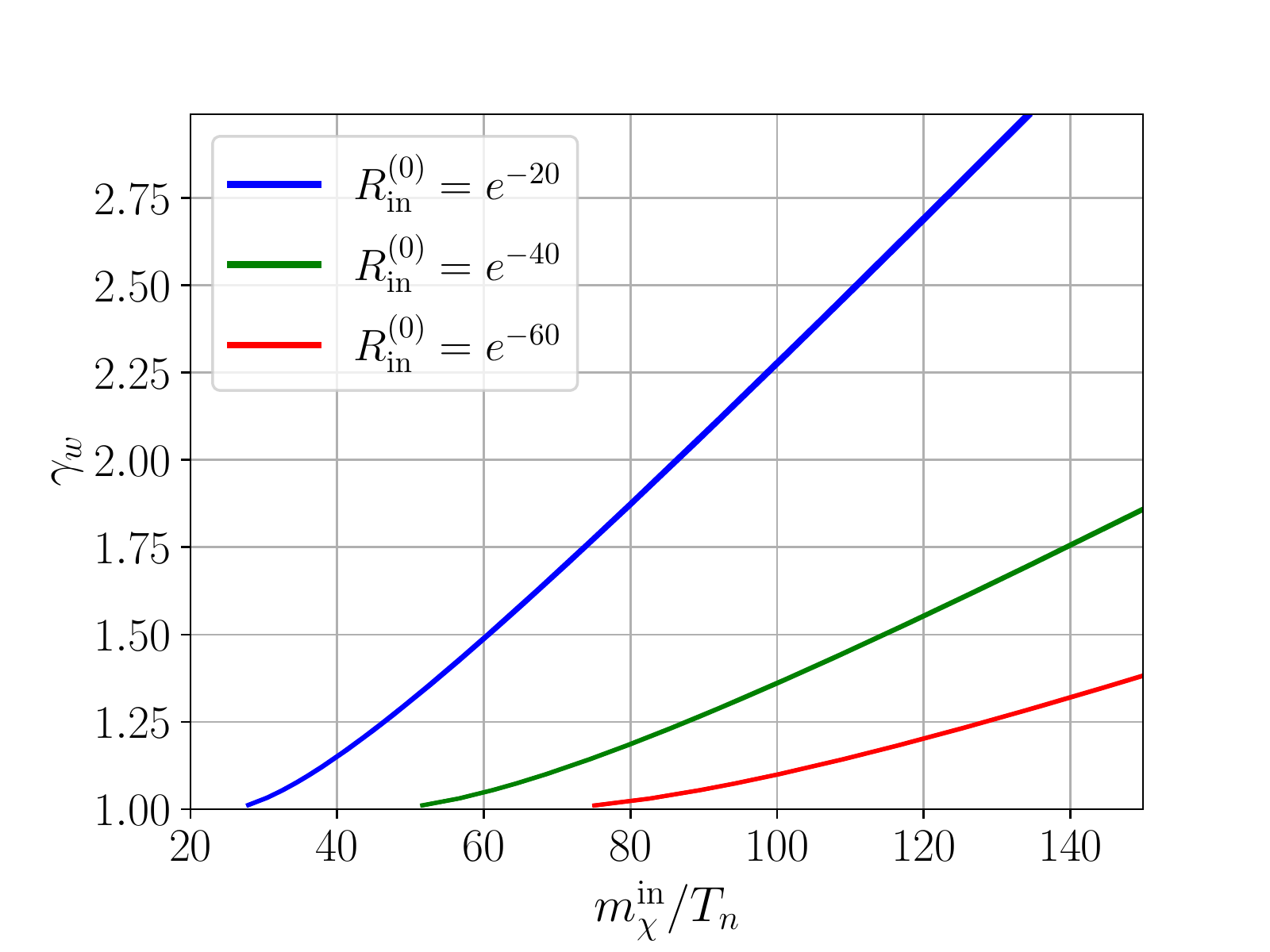}
	\end{minipage}}	
	\caption{Left: contour plot of $\gamma_w$ and $m_\chi^{\mathrm{in}}/T_n$ with DM relic density $\Omega_{\mathrm{DM}} ^{(0)} h^2 =0.12$ without hydrodynamic effects. Right: penetration rate $R_{\rm in}^{(0)}$ as functions of $\gamma_w$ and $m_\chi^{\mathrm{in}}/T_n$ without hydrodynamic effects.}\label{filter}.
\end{figure}

In the original work of Ref.~\cite{Baker:2019ndr} they do not consider the variation of velocity and temperature across the bubble wall from the hydrodynamic effects, namely, they choose 
\begin{equation}\label{nohy}
\boxed{\tilde v_{\mathrm{pl}} = v_w, \quad T=T'=T_n \quad \rightarrow \quad \Omega_{\mathrm{DM}}^{(0)} h^2 ~[2]\,\,.}
\end{equation}
We show the contour plot of the observed DM relic density $\Omega_{\mathrm{DM}}^{(0)}h^2=0.12$ in the left panel of Fig.~\ref{filter}. It is shown that in order to fulfill the observed DM relic abundance we cannot generally impose an ultrarelativistic bubble wall with $\gamma_w \gg 1$. For $v_w \rightarrow 0.01$, i.e., $\gamma_w \rightarrow 1$, we have to impose $m_\chi^{\rm in}/T_n$ around 40. And for $v_w > 0.9$, i.e., $\gamma_w > 2.3$, $ m_\chi^{\rm in}/T_n > 100$ is generally needed to efficiently filter the DM.

One can define the DM penetration rate $R_{\rm in}^{(0)}$ as in the original works as
\begin{eqnarray}
R_{\rm in}^{(0)}=\frac{n_{\chi}^{\text {in }}}{n_{\chi}^{\text {out }}}\,\,,
\end{eqnarray}
where $n_{\chi}^{\text {out }}=g_\chi T^3 / \pi^2$ is the equilibrium DM number density outside the bubble wall at the time of the phase transition. The penetration rate $R_{\rm in}^{(0)}$ is shown in the right panel of Fig.~\ref{filter}. This quantity evaluates the filtering ability of the bubble wall. The larger the $R_{\rm in}^{(0)}$ is, the weaker the filtering effect is.


Actually, the temperature and velocity are not constants across the bubble wall. As the bubble wall expands, the vacuum energy releases and then causes reheating and bulk motions of the plasma around the bubble wall. We denote the quantities in front of the bubble wall with $+$ and behind the wall with $-$. Then in the filtered DM mechanism, $\tilde v_{\rm pl}$ and $T$ in Eq.~\eqref{feq} should be evaluated in front of the bubble wall and $T'$ is the temperature behind the bubble wall,
\begin{equation}\label{hyc}
\boxed{\quad \quad \quad \tilde v_{\rm pl}=\tilde v_+, \quad T=T_+, \quad T'=T_-   \qquad \text{(this work with hydrodynamic effects)}\,\,.}
\end{equation}
Then Eq.~\eqref{nana1} should be rewritten as
\begin{eqnarray}\label{nana2}
n_\chi^{\text {in }} \simeq \frac{g_{\chi} T_{+}^3}{\gamma_w v_w}\left(\frac{\tilde \gamma_{+}\left(1- \tilde v_{+}\right) m_\chi^{\mathrm{in}} / T_{+}+1}{4 \pi^2 \tilde \gamma_{+}^3\left(1- \tilde v_{+}\right)^2}\right) e^{-\frac{\tilde \gamma_{+}\left(1- \tilde v_{+}\right) m_\chi^{\mathrm{in}}}{T_{+}}} \,\,,
\end{eqnarray}
and we have a DM relic density of
\begin{eqnarray}\label{oh2hy}
\Omega_{\mathrm{DM}}^{(\mathrm{hy})} h^2 =\frac{m_{\chi}^{\mathrm{in}} (n_\chi^{\rm in}+n_{\bar\chi}^{\rm in})}{\rho_c / h^2} \frac{ g_{\star  0} T_0^3}{ g_{\star }\left(T_{-}\right) T_{-}^3}\simeq 6.29 \times 10^8 \frac{m_{\chi}^{\mathrm{in}}}{\mathrm{GeV}} \frac{(n_\chi^{\rm in}+n_{\bar\chi}^{\rm in})}{g_{\star }\left(T_{-}\right) T_{-}^3}\,\,.
\end{eqnarray}

This can also be expressed in terms of the renewed penetration rate $R_{\rm in}^{(\mathrm{hy})}$,
\begin{eqnarray}
	\Omega_{\mathrm{DM}}^{(\mathrm{hy})} h^2 =  6.37 \times 10^7 \frac{m_{\chi}^{\mathrm{in}}}{\mathrm{GeV}} \frac{2g_\chi}{g_{\star }(T_-)} \left(\frac{T_+}{T_-}\right)^3 R_{\rm in}^{(\mathrm{hy})} \,\,.
\end{eqnarray}
with 
\begin{eqnarray}\label{Rinana}
	R_{\rm in}^{(\mathrm{hy})} = \frac{1}{\gamma_w v_w}\left(\frac{\tilde \gamma_{+}\left(1- \tilde v_{+}\right) m_\chi^{\mathrm{in}} / T_{+}+1}{4 \tilde \gamma_{+}^3\left(1- \tilde v_{+}\right)^2}\right) e^{-\frac{\tilde \gamma_{+}\left(1- \tilde v_{+}\right) m_\chi^{\mathrm{in}}}{T_{+}}}\,\,.
\end{eqnarray}

There are generally two distinct cases: 1) For a phase transition with $\alpha_n \lesssim 1$, we expect that $T_- $ will be different with $T_+$ by a ${\cal O}(1)$ factor and the DM relic abundance is dominantly set by filtering effects. 2) However, for strong supercooling phase transition $\alpha_n \gg 1$, we expect the wall velocity is ultrarelativistic and the dominant mechanism is \emph{super-cool DM}~\cite{Hambye:2018qjv} which produce very large $T_-/T_+$ to suppress the relic density of DM. In the latter case, one may still want to quantify the filtering effects in order to get accurate results. An accurate evaluation of $R_{\rm in}^{(\mathrm{hy})}$ is important in a given model.

It is obviously important to evaluate the temperature and velocity in front of and behind the bubble wall. On one hand, as the typical energy of DM in front of the bubble wall is proportional to the temperature $T_+$, higher temperature $T_+$ in front of the bubble wall means stronger ability of DM to pass through the bubble wall. However, a larger $T_-$ may lead to a larger vacuum value behind the wall, producing a larger $m_\chi^{\mathrm{in}}$. This makes it harder for DM to pass through the bubble wall. On the other hand, the velocity determines the momentum distribution of DM, which affects the DM number density. The hydrodynamic effects are first mentioned by Ref.~\cite{Chway:2019kft} without calculations. In this work we demonstrate how to calculate the hydrodynamic effects in details.

\section {Phase transition dynamics}\label{PTdynamics}
\subsection{Toy model}
To work in the filtered DM mechanism, we assume a simple DM model
\begin{eqnarray}
	\mathscr{L}=- y_\chi \overline{\chi} \Phi \chi-V_{}(\Phi) -\kappa\Phi^{\dagger}\Phi H^{\dagger}H + H.c.\,\,,
\end{eqnarray}
where $\Phi$ is a scalar doublet and $V(\Phi)$ is the tree-level potential of $\Phi$.  $\chi$ is a Dirac particle which would be the DM candidate. This model can represent a variety of phase transition models like the inert singlet, inert doublet models. The background field is defined as
\begin{eqnarray}
	\Phi = (0, \phi/\sqrt{2})^T \,\,.
\end{eqnarray}
The mixing term $\kappa\Phi^{\dagger}\Phi H^{\dagger}H$ behaves as a portal with which the DM particles can become thermal equilibrium with standard model (SM) plasma outside the bubble wall then they share the same temperature and velocity. And due to this portal coupling $\phi$ can annihilate away after phase transition.
We assume the effective scalar potential at finite temperature is of the form
\begin{eqnarray}
V_{\mathrm{eff}}(\phi, T)=\frac{\mu^2+D T^2}{2} \phi^2- C T \phi^3+\frac{\lambda}{4} \phi^4 - \frac{g_\star\pi^2T^4}{90}\,\,,
\end{eqnarray}
where $g_\star$ is the effective number of freedom at phase transition. We choose $g_\star=120$ for simplicity as counting some extra particle species.

The critical temperature at which $V_{\mathrm{eff}}(0, T_c)=V_{\mathrm{eff}}(\phi_c, T_c)$ is
\begin{eqnarray}
T_c=\frac{2}{\lambda D-2 C^2}\left[\frac{\sqrt{\lambda D\left(\lambda D-2 C^2\right) T_b^2}}{2}\right]\,\,,
\end{eqnarray}
where
\begin{eqnarray}
T_b^2=-\frac{\mu^2}{D}=\frac{\lambda}{D} v_{0}^2\,\,,
\end{eqnarray}
is the temperature when the potential barrier vanishes. We choose zero-temperature vacuum value $v_{0}=246~\rm{GeV}$ for simplicity. The two minima are
\begin{eqnarray}\label{vevT}
\langle \phi \rangle=0, \quad \frac{3 C T}{2 \lambda}\left[1+\sqrt{1-\frac{4 \lambda\left(\mu^2+D T^2\right)}{9 C^2 T^2}}\right]\,\,.
\end{eqnarray}

Then the DM mass deep inside the bubble is given by,
\begin{eqnarray}
	m_\chi^{\mathrm{in}} = \frac{y_\chi \phi_-}{\sqrt{2}}\,\,,
\end{eqnarray}
where $\phi_- = \phi(T_-)$ is the field value that is evaluated at the temperature behind the bubble wall. $\phi_n$ is the field value at nucleation temperature $T_n$.

We use \emph{CosmoTransitions}~\cite{Wainwright:2011kj} to find the phase transition points. The bubbles begin to  nucleate at the temperature $T_n$ when bounce action $S_3(T_n)/T_n\simeq 142$. Actually, this simple potential has semi-analytical approximation for the tunneling action~\cite{Adams:1993zs,Ellis:2020awk}
\begin{eqnarray}\label{Sana}
	\frac{S_3}{T}=\frac{C}{\lambda^{3/2}} \frac{64\pi \sqrt{\delta}(\beta_1 \delta +\beta_2 \delta^2 +\beta_3\delta^3)}{81(2-\delta)^2}
\end{eqnarray}
with
\begin{eqnarray}\label{delta}
	\delta = \frac{\lambda (\mu^2+DT^2)}{C^2 T^2}
\end{eqnarray}
and $\beta_1=8.2938$, $\beta_2=-5.5330$, $\beta_3=0.8180$.
 
\begin{figure}[htbp]
	\begin{center}
		\includegraphics[width=1.25\linewidth]{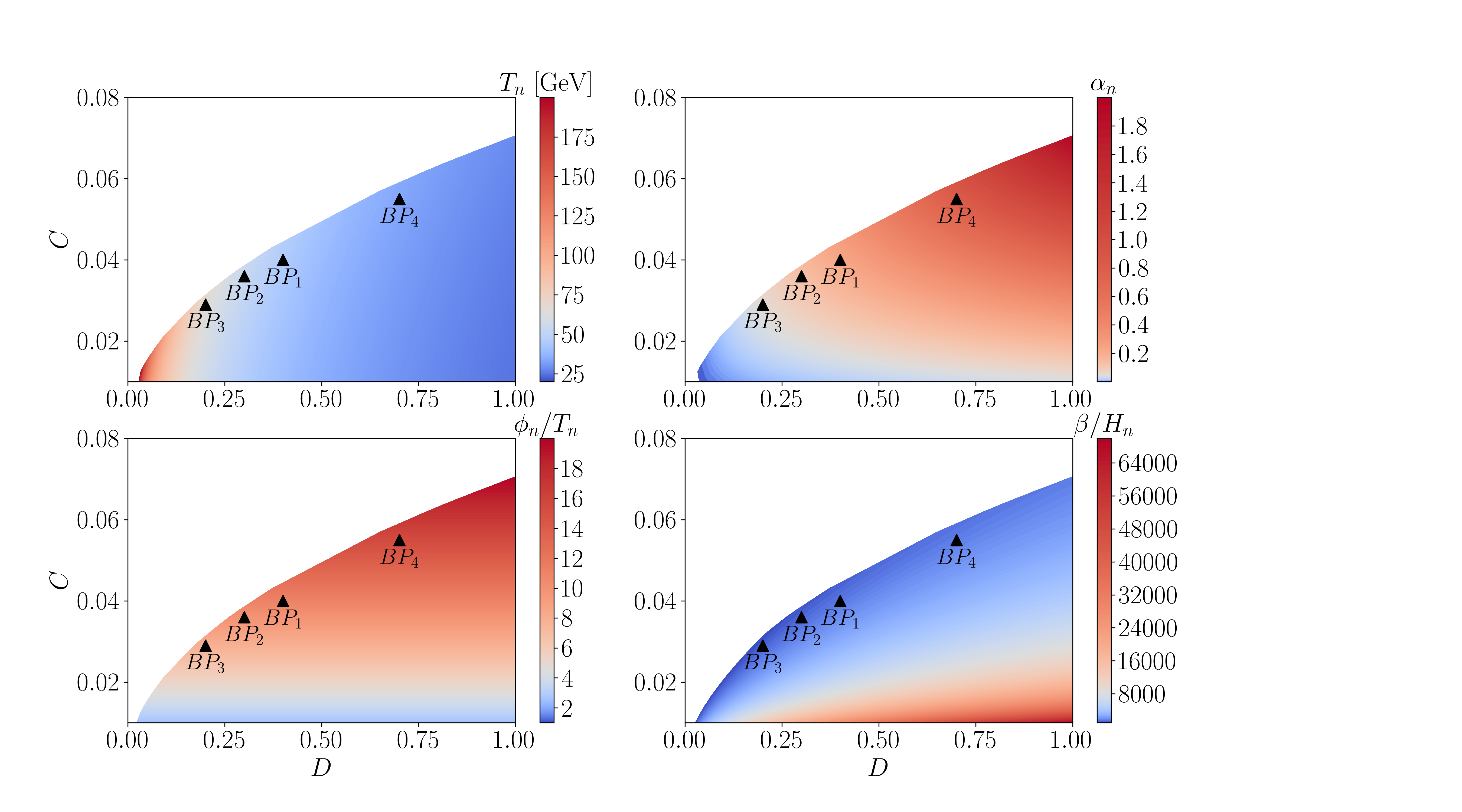}
		\caption{The four phase-transition parameters $T_n$, $\alpha_n$, $\phi_n/T_n$ and $\beta/H_n$ in the toy model with fixed $\lambda=0.01$. }
		\label{scan}
	\end{center}
\end{figure}

Then we use the following definition of  the phase transition strength:
 \begin{eqnarray}\label{PTstrength}
 	\alpha \equiv \frac{\left(1-\frac{T}{4} \frac{\partial}{\partial T}\right) \Delta V_{\mathrm{eff}}}{\pi^2 g_{\star} T^4 / 30}\,\,,
 \end{eqnarray}
where $\Delta V_{\mathrm{eff}}$ is the potential difference between the false and true vacuum.
Then the inverse duration of the phase transition is
\begin{eqnarray}
	\beta = -\frac{d}{dt}\frac{S_3(T)}{T} = H(T)T\frac{d}{dT}\frac{S_3(T)}{T} \,\,.
\end{eqnarray}
We can also have the analytic expression of $\alpha$ and $\beta$, which is shown in Appendix~\ref{PTa}. In Appendix~\ref{PTa} it is also shown that the analytic expression and CosmoTransitions give almost the same results. We show four benchmark points in Table.~\ref{ptable}. In our work we mainly work in $BP_1$ and we will also show the results of other three benchmark points. We use the semi-analytic method to scan the parameters in toy model as shown in Fig.~\ref{scan}. We fix $\lambda =0.01$ and then show four phase-transition parameters as functions of $D$, $C$ in the toy model: the nucleation temperature $T_n$, the phase transition strength $\alpha_n$, the wash out parameter $\phi_n/T_n$ and phase transition rate $\beta/H_n$, where $H_n=H(T_n)$ is the Hubble constant at nucleation.

\begin{table}[t]
	\centering
	
	\setlength{\tabcolsep}{3mm}
	
	\begin{tabular}{c|c|c|c|c|c|c|c}
		\hline\hline
		& $D$ & $C$ & $\lambda$ &$T_n $~[GeV] &$\alpha_n$ & $\beta/H_n$ &$\phi_n/T_n$\\
		\hline
		$BP_1$	&0.4 &0.04  &0.01 & 45.13& 0.23 & 1531.69 & 11\\
		\hline
		$BP_2$	& 0.3 & 0.036 & 0.01 & 53.42 & 0.13 & 1299.67 & 10\\
		\hline
		$BP_3$	& 0.5 & 0.04 &0.01 & 39.03& 0.31 & 2044.65 & 11.1\\
		\hline 
		$BP_4$	& 0.7 & 0.055  & 0.01 & 33.6 & 0.81  & 1659.02 & 15.4\\
		\hline\hline
	\end{tabular}
	\caption{Four sets of benchmark parameters and the corresponding phase transition parameters.}\label{ptable}
\end{table}

\subsection{Bubble wall velocity}
Precise calculation of bubble wall velocity is still difficult now. In reality, this should be done by evaluating the interactions between the bubble wall and the plasma. The friction comes from the non-equilibrium part of heavy particles that have masses comparable with the phase transition temperature. The solution of bubble wall velocity requires us to solve the EOM including non-equilibrium part~\cite{Moore:1995si,Moore:1995ua,Jiang:2022btc}
\begin{eqnarray}
	\partial^2\phi+\frac{\partial V_{\text {eff }}}{\partial \phi}=\sum_{i=\mathrm{B}, \mathrm{F}} g_i \frac{\mathrm{d} m_i^2}{\mathrm{~d} \phi} \int \frac{\mathrm{d}^3 p}{(2 \pi)^3} \frac{\delta f_i}{2 E_i}\,\,.
\end{eqnarray}
However, even in local equilibrium, there is still an effective friction on the bubble wall~\cite{BarrosoMancha:2020fay,Ai:2021kak,Balaji:2020yrx}. The general quantum field theoretic formula shows that~\cite{BarrosoMancha:2020fay}
\begin{eqnarray}\label{le}
	\Delta P_{\rm le} = (\gamma_w^2-1)T_n\Delta s\,\,.
\end{eqnarray}
with $\Delta s$ is the difference of the entropy density between the true and the false vacuum.

For ultra-relativistic bubble wall with $\gamma_w T_n\gg m_i$ the leading-order contribution of friction is~\cite{Bodeker:2009qy}
\begin{eqnarray}
	\Delta P_{\rm LO} = \sum_i g_i c_i \frac{\Delta m_i^2}{24} T_n^2\,\,,
\end{eqnarray}
where $g_i$ is the number of freedom of the particles and $\Delta m_i^2 = m_i^2(\phi_-)-m_i^2(\phi_+)$ with $\phi_-$ and $\phi_+$ are the vacuum value behind and in front of the bubble wall, respectively. $c_i=1$ for bosons and $c_i = 1/2$ for fermions. The next-to-leading-order (NLO) friction is still being debated.
One is proportional to $\gamma_w$~\cite{Gouttenoire:2021kjv}
\begin{eqnarray}
	\Delta P^{(1)}_{\rm NLO} \approx \gamma_w\sum_{j\in V} \lambda_j m_j T_n^3 \text{log}\frac{m_j}{\mu_{ref}} \,\,,
\end{eqnarray}	
where $\sum_{j \in V} $ sums over only gauge bosons and $\lambda_j$ is their gauge couplings. $\mu_{ref} \approx \lambda_j T_n$ is an IR cutoff. However, Ref.~\cite{Hoche:2020ysm} reports another expression for the NLO friction which is proportional to $\gamma_w^2$:
\begin{eqnarray}
	\Delta P^{(2)}_{\rm NLO} \approx \gamma_w^2 \sum_{j\in V}\lambda_j g_j T_n^4 \,\,.
\end{eqnarray}

In Ref.~\cite{Chway:2019kft}, they pointed that, for a filtered DM mechanism, the main friction should come from the heavy particles :
\begin{eqnarray}\label{pheavy}
	\Delta P_{\rm heavy}=\frac{d_n g_{\star} \pi^2}{90}\left(1+v_w\right)^3 \gamma_\omega^2 T_{n}^4 \,\,,
\end{eqnarray}
where
\begin{eqnarray}
	d_n \equiv \frac{1}{g_{\star}}\left[\sum_{0.2 m_i>\gamma_w T_{n}}\left(g_i^b+\frac{7}{8} g_i^f\right)\right]\,\,,
\end{eqnarray}
with $g_i^b$ and $g_i^f$ as the number of degrees of freedom of the bosons and fermions, respectively. This has a similar form as Eq.~\eqref{le}. For heavy particles in a false vacuum they have an entropy density $s \propto T^3$, and in true vacuum their contribution to entropy density is approximately zero.

In this work, we focus on the hydrodynamic effects on the DM relic density. And the bubble wall velocity is an important parameter to determine the hydrodynamic mode. To obtain more general results, 
we do not specify the definite bubble wall velocity, but discuss three hydrodynamic modes for different bubble wall 
velocities.

\section{Velocity and temperature distribution from hydrodynamic effects with zero-width bubble wall}\label{hydro}
As the bubble wall is expanding, the phase transition happens and the latent heat is injected, which causes reheating and bulk motions of the plasma. This produces the temperature and velocity profiles around the bubble wall. In this section we discuss the hydrodynamic effects without considering the thickness of the bubble wall~\cite{Espinosa:2010hh}.

\subsection{Continuity equations}
The energy-momentum tensor (EMT) for the scalar field $\phi$ is
\begin{eqnarray}
	T_\phi^{\mu \nu}=\partial^\mu \phi \partial^\nu \phi-g^{\mu \nu}\left[\frac{1}{2} (\partial \phi)^2 -V_{T=0}\left(\phi\right)\right]\,\,,
\end{eqnarray}
where $V_{T=0}\left(\phi\right)$ is the effective potential at zero temperature that includes one-loop quantum corrections.

The energy-momentum tensor for plasma is
\begin{eqnarray}
	T_{\mathrm{pl}}^{\mu \nu}=\sum_i \int \frac{d^3 k}{(2 \pi)^3 E_i} k^\mu k^\nu f_i^{\mathrm{eq}}\left(k\right)\,\,,
\end{eqnarray}
which can be parametrized as a perfect fluid
\begin{eqnarray}
	T_{\mathrm{pl}}^{\mu \nu}=\omega_{\mathrm{pl}} u_{\mathrm{pl}}^\mu u_{\mathrm{pl}}^\nu-p_{\mathrm{pl}} g^{\mu \nu}\,\,,
\end{eqnarray}
where $u_{\mathrm{pl}}^{\mu} = (\gamma,\gamma \vec{v})$ with $\gamma = 1/\sqrt{1-v^2}$ in the frame of the bubble center, the thermal pressure $p_{\mathrm{pl}}$ and enthalpy $\omega_{\mathrm{pl}}$ are
\begin{eqnarray}
	\begin{aligned}
		p_{\mathrm{pl}} & =\pm T \sum_i \int \frac{d^3 k}{(2 \pi)^3} \log \left[1 \pm \exp \left(-E_i / T\right)\right]\,\,, \\
		\omega_{\mathrm{pl}} & =T \frac{\partial p_{\mathrm{pl}}}{\partial T} = e_{\mathrm{pl}}+p_{\mathrm{pl}}
	\end{aligned}
\end{eqnarray}
with $E_i = \sqrt{k^2+m_i^2}$ defined in the plasma rest frame. The signs $+/-$ denote fermions/bosons.

Because for scalar field EMT $\omega_{\phi}=e_{\phi}+p_{\phi}=0$ and $p_{\phi}=-V_{T=0}(\phi)$, we can write the EMT of the total system as 
\begin{eqnarray}
	T_{\mathrm{fl}}^{\mu \nu} =T_{\phi}^{\mu \nu} + T_{\mathrm{pl}}^{\mu \nu}= \omega u^{\mu} u^{\nu} - pg^{\mu \nu} 
	\,\,,
\end{eqnarray}
with $\omega = \omega_{\mathrm{pl}}$, $u^{\mu}=u_{\mathrm{pl}}^{\mu}$, and $p=p_{\mathrm{pl}} - V_{T=0}$.

The equation of state can be parametrized by using the bag model. In the false vacuum,
\begin{eqnarray}\label{plus}
p_{+}=\frac{1}{3} a_{+} T_{+}^4-\epsilon, \quad e_{+}=a_{+} T_{+}^4+\epsilon \,\,,
\end{eqnarray}
and in the broken phase
\begin{eqnarray}\label{minus}
p_{-}=\frac{1}{3} a_{-} T_{-}^4, \quad e_{-}=a_{-} T_{-}^4 \,\,.
\end{eqnarray}
Here
\begin{eqnarray}
\left.a_{ \pm} \equiv \frac{3}{4 T_{ \pm}^3} \frac{\partial p}{\partial T}\right|_{ \pm}=\frac{3 \omega_{ \pm}}{4 T_{ \pm}^4}, \quad \epsilon_{ \pm}=\frac{1}{4}\left(e_{ \pm}-3 p_{ \pm}\right) \,\,,
\end{eqnarray}
where $a_{\pm}$ are the number of freedom in the false and true vacuum respectively. Mostly $a_{+}>a_{-}$ as counting the heavy particles across the bubble wall. $\epsilon$ denotes the false-vacuum energy, which is often defined to be zero in the broken, true-minimum phase.

We consider the continuity equations that come from the energy-momentum conservation,
\begin{eqnarray}
	\nabla_{\mu} T^{\mu \nu}=u^\nu \nabla_\mu\left(u^\mu \omega\right)+u^\mu \omega \nabla_\mu u^\nu-\nabla^\nu p=0 \,\,.
\end{eqnarray}
We project the direction of the flow of the fluid, and using $u_\mu \nabla_\nu u^\mu=0$, we get
\begin{eqnarray}\label{parallel}
\nabla_\mu\left(u^\mu \omega\right)-u_\mu \nabla^\mu p=0 \,\,.
\end{eqnarray}
We next project perpendicular to the flow with some space-like vector $\bar{u}=$ $\gamma(v, \mathbf{v} / v)$ such that $\bar{u}_\mu u^\mu=0, \bar{u}^2=-1$. Then we have
\begin{eqnarray}\label{perpend}
\bar{u}^\nu u^\mu \omega \nabla_\mu u_\nu-\bar{u}^\nu \nabla_\nu p=0 \,\,.
\end{eqnarray}
Since there is no characteristic distance scale in the problem, the solution could be described by a self-similar parameter $\xi = r/t$, where $r$ is the distance from the bubble center and $t$ the time from nucleation. Then the gradients will be turned into 
\begin{eqnarray}
	u_\mu \nabla^\mu=-\frac{\gamma}{t}(\xi-v) \partial_{\xi}, \quad \bar{u}_\mu \nabla^\mu=\frac{\gamma}{t}(1-\xi v) \partial_{\xi} \,\,.
\end{eqnarray}
By using
\begin{eqnarray}
	\nabla_{\mu} u^{\mu} = \frac{jv}{\xi}\frac{\gamma}{t}+\frac{\gamma}{t}\gamma^2(1-\xi v)\partial_{\xi}v \,\,,
\end{eqnarray}
with $j=0,1,2$ for planar, cylindrical, and spherical cases, respectively.
Then Eqs.~\eqref{parallel} and \eqref{perpend} can be transformed into
\begin{eqnarray}\label{continuity}
	\begin{aligned}
		(\xi-v) \frac{\partial_{\xi} e}{\omega} & =j \frac{v}{\xi}+\left[1-\gamma^2 v(\xi-v)\right] \partial_{\xi} v \,\,,\\
		(1-v \xi) \frac{\partial_{\xi} p}{\omega} & =\gamma^2(\xi-v) \partial_{\xi} v \,\,.
	\end{aligned}
\end{eqnarray}

The derivatives $\partial_{\xi} e$ and $\partial_{\xi} p$ can be related through the speed of sound in the plasma, $c_s^2 \equiv(d p / d T) /(d e / d T)$, then from Eq.~\eqref{continuity} we have
\begin{equation}\label{velocity continuity}
\begin{aligned}
j \frac{v}{\xi} &=\gamma^2(1-v \xi)\left[\frac{\mu^2}{c_s^2}-1\right] \partial_{\xi} v \,\,, \\
\frac{\partial_{\xi} \omega}{\omega} &= \left(1+\frac{1}{c_s^2}\right) \gamma^2 \mu \partial_{\xi} v \,\,.
\end{aligned}
\end{equation}
with $\mu$ as the Lorentz-transformed fluid velocity,
\begin{eqnarray}
\mu(\xi, v)=\frac{\xi-v}{1-\xi v} \,\,.
\end{eqnarray}
In many cases $c_s^2$ only slightly deviates from $1 / 3$.
After getting $v(\xi)$, we can integrate Eq.~\eqref{velocity continuity} to get
\begin{eqnarray}\label{enthalpy}
\omega(\xi)=\omega_0 \exp \left[\int_{v_0}^{v(\xi)}\left(1+\frac{1}{c_s^2}\right) \gamma^2 \mu d v\right]\,\,.
\end{eqnarray}

\subsection{Matching conditions}
In order to solve the continuity equations, we must impose some boundary conditions. There are two boundaries in our cases, the bubble wall at $\xi_w=v_w$ and the shock front $\xi_{sh}$. In the reference frame of the bubble wall, we can again use the energy-momentum conservation across the boundaries, remember that we denote $\tilde v$ as velocity in the frame of the bubble wall,
\begin{eqnarray}\label{flow}
\omega_{+} \tilde v_{+}^2 \tilde\gamma_{+}^2+p_{+}=\omega_{-} \tilde v_{-}^2 \tilde \gamma_{-}^2+p_{-}, \quad \omega_{+} \tilde v_{+} \tilde \gamma_{+}^2=\omega_{-} \tilde v_{-} \tilde \gamma_{-}^2 \,\,.
\end{eqnarray}
From these equations one obtains the relations
\begin{eqnarray}\label{vpm1}
\tilde v_{+} \tilde v_{-}=\frac{p_{+}-p_{-}}{e_{+}-e_{-}}=\frac{1-\left(1-3 \alpha_{+}\right) r_{\omega}}{3-3\left(1+\alpha_{+}\right) r_{\omega}}, \quad \frac{\tilde v_{+}}{\tilde v_{-}}=\frac{e_{-}+p_{+}}{e_{+}+p_{-}}=\frac{3+\left(1-3 \alpha_{+}\right) r_{\omega}}{1+3\left(1+\alpha_{+}\right) r_{\omega}} \,\,.
\end{eqnarray}
with $\alpha_{+} \equiv \epsilon /\left(a_{+} T_{+}^4\right)$ and $r_{\omega}=\omega_+/\omega_-=(a_+T_+^4)/(a_-T_-^4)$. The fluid velocities on each side in the bubble center frame are given by $v_{\pm} = \mu(\xi_w,\tilde v_{\pm})$.


Substituting Eqs.~\eqref{plus} and \eqref{minus} into Eq.~\eqref{vpm1}, we can solve for $\tilde v_+$ as function of $\tilde v_-$,
\begin{eqnarray}\label{tvptvm}
	\tilde v_{+}=\frac{1}{1+\alpha_{+}}\left[\left(\frac{\tilde v_{-}}{2}+\frac{1}{6 \tilde v_{-}}\right) \pm \sqrt{\left(\frac{\tilde v_{-}}{2}+\frac{1}{6 \tilde v_{-}}\right)^2+\alpha_{+}^2+\frac{2}{3} \alpha_{+}-\frac{1}{3}}\right]\,\,.
\end{eqnarray}

\begin{figure}[htbp]
	\begin{center}
		\includegraphics[width=0.7\linewidth]{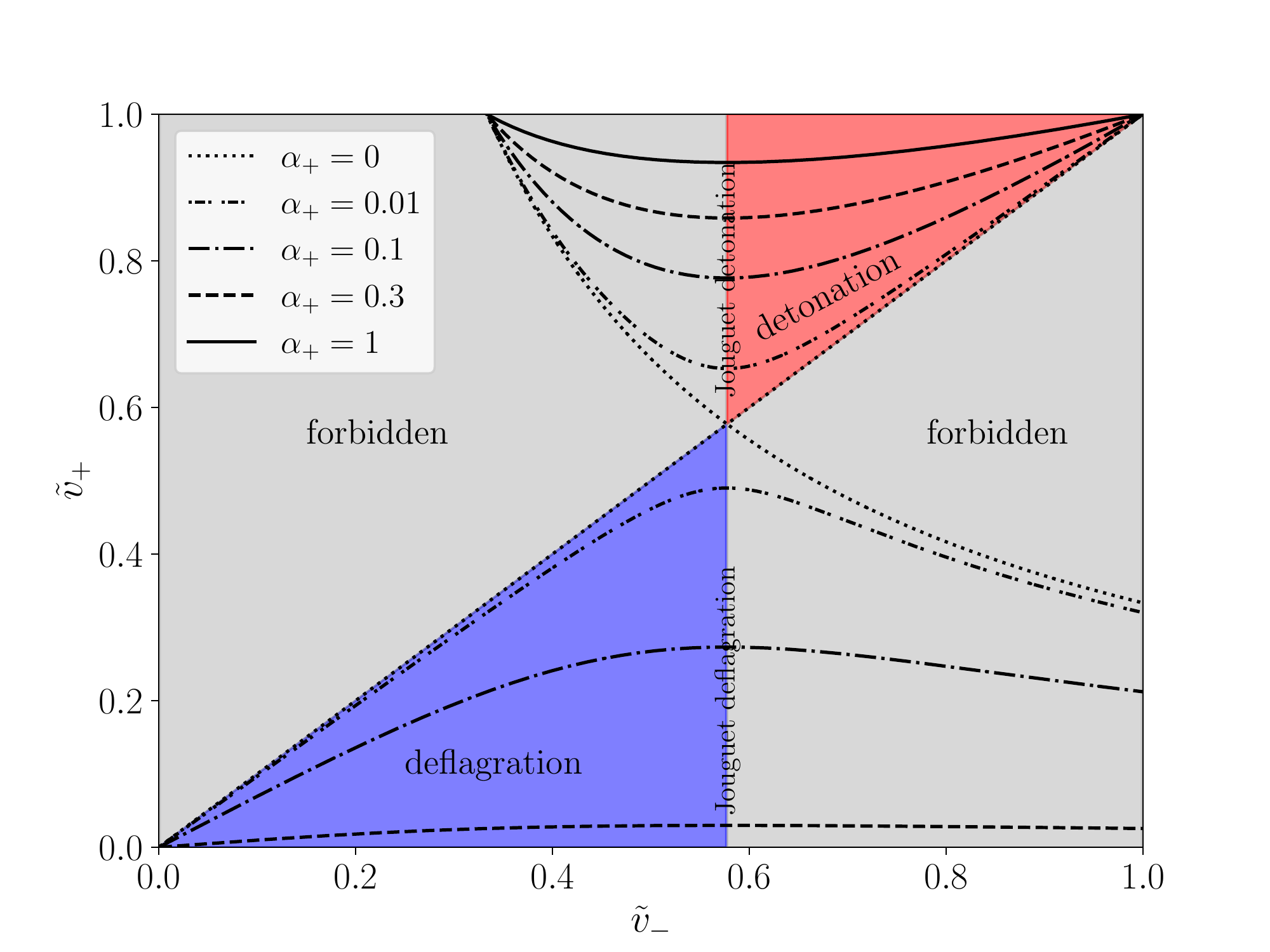}
		\caption{Velocities in front of and behind the bubble wall. The upper branch and the lower branch correspond to detonation and deflagration respectively. The gray areas are forbidden because they are not physical solutions.}
		\label{vpvm}
	\end{center}
\end{figure}

We show the two branches of solutions in Fig.~\ref{vpvm}. The upper branch corresponds to the detonation that the incoming flow is supersonic ($\tilde v_+ > c_s$) and is faster than outgoing flow ($\tilde v_+>\tilde v_-$). For deflagration $\tilde v_+<c_s$ and $\tilde v_+<\tilde v_-$. One can see that there are no deflagration solutions for $\alpha_+>1/3$. The process with $\tilde v_-=c_s$ is called a Jouguet detonation. We then have
\begin{eqnarray}\label{Jouguet velocity}
	\tilde v_+ = v_J^{}(\alpha_+) = \frac{1 + \sqrt{\alpha_+(2+3\alpha_+)}}{\sqrt{3}(1+\alpha_+)} \,\,.
\end{eqnarray} 
For deflagration mode, in order to satisfy the condition that $v(\xi \rightarrow \infty) = 0$, there should be a shock front at $\xi_{sh}$ in front of the bubble wall. We shall use index 1 for variables behind the shock front and index 2 for variables in front of the shock front. Energy-momentum conservation across the shock front gives us a similar form as Eq.~\eqref{vpm1},
\begin{eqnarray}\label{shock}
\tilde v_1 \tilde v_2 = \frac{1}{3}, \quad \frac{\tilde v_1}{\tilde v_2}=\frac{3T_2^4+T_1^4}{3T_1^4+T_2^4} \,\,. 
\end{eqnarray}

\begin{figure}[htbp]
	\centering
	\subfigure{
		\begin{minipage}[t]{0.5\linewidth}
			\centering
			\includegraphics[scale=0.55]{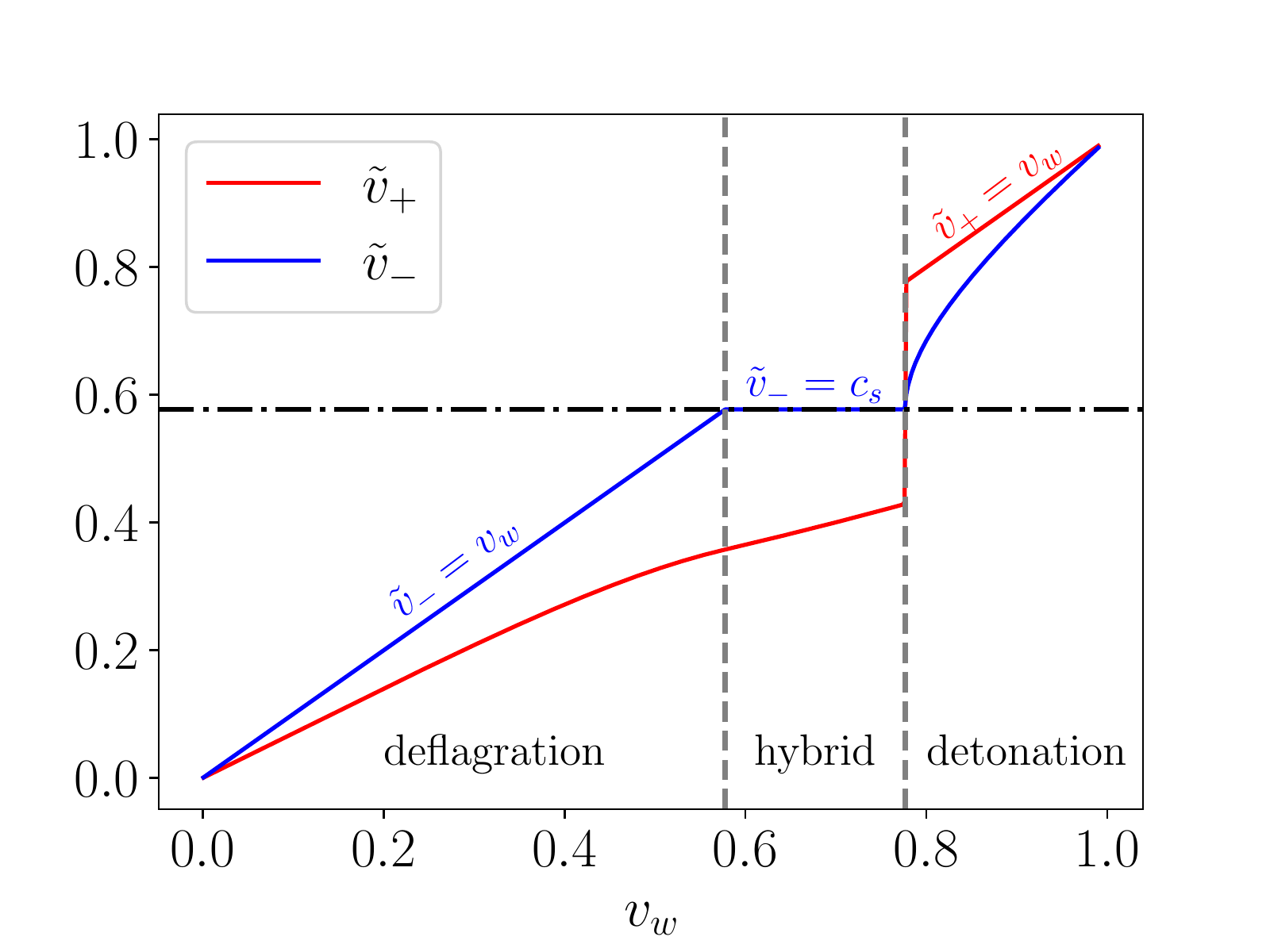}
	\end{minipage}}%
	\subfigure{
		\begin{minipage}[t]{0.5\linewidth}
			\centering
			\includegraphics[scale=0.55]{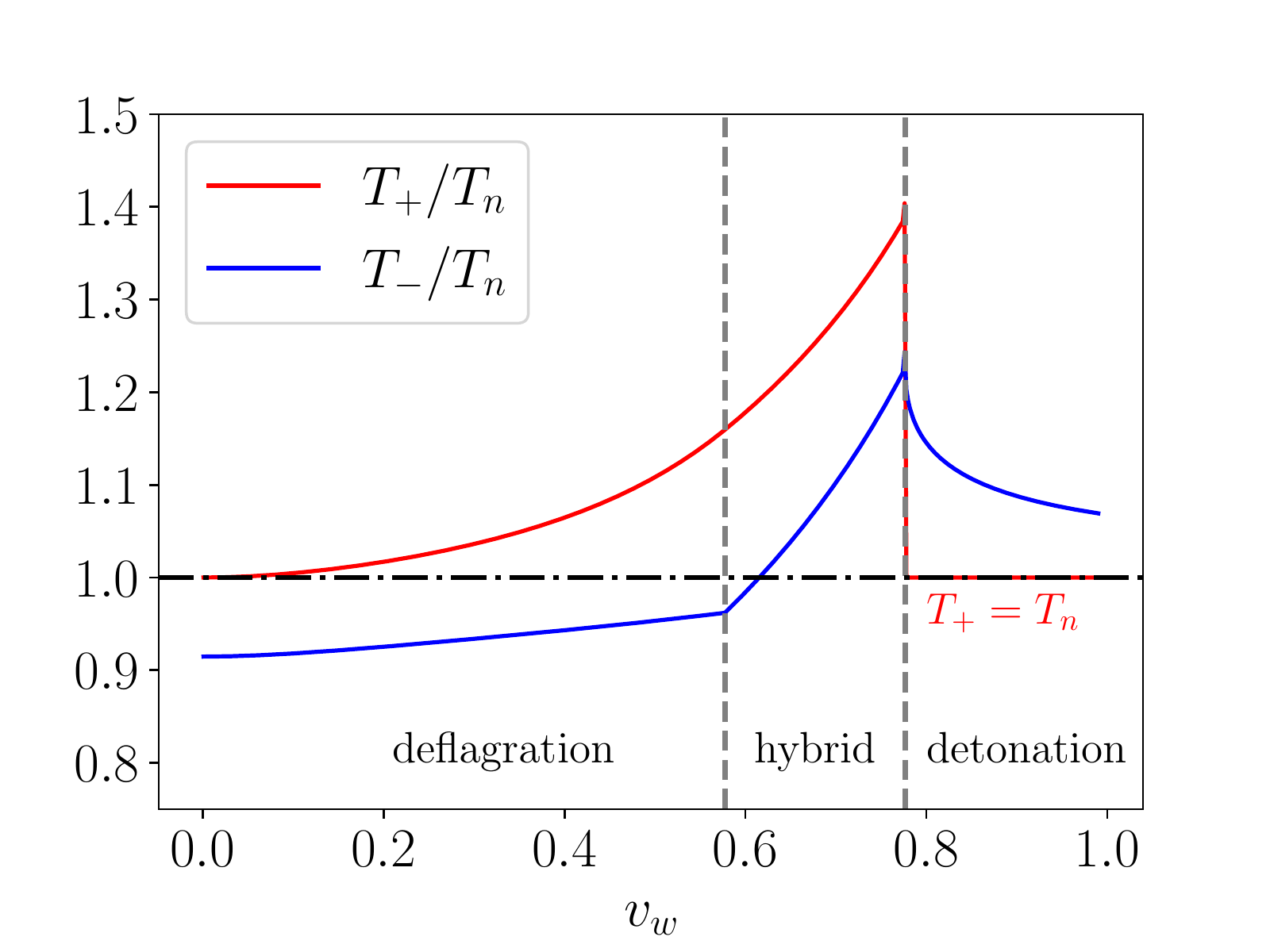}
	\end{minipage}}	
	\caption{Left: flow velocity in the bubble wall frame $\tilde v_+$ and $\tilde v_-$ as functions of $v_w$ for $\alpha_n=0.1$ for spherical bubbles. Right: temperature in front of and behind the bubble wall as functions of $v_w$ for $\alpha_n=0.1$  for spherical bubbles.}\label{vTvw}.
\end{figure}

\subsection{Spherical approximation}\label{sa}
In the spherical case, for detonation the velocity in front of the bubble wall $v_+=0$ in the bubble center frame and $T_+=T_n$. Then we have $\tilde v_+= \xi_w = v_w$ and $\alpha_+=\alpha_n$ where $\alpha_n=\epsilon/(a_+T_n^4)$. The bubble wall is followed by a rarefaction wave that ends at $\xi=c_s$. The rarefaction wave is depicted by Eq.~\eqref{velocity continuity}. The boundary condition is
\begin{eqnarray}
	v(\xi_w)=v_-=\mu(\xi_w,\tilde v_-) \,\,,
\end{eqnarray}
with $\tilde v_-$ given by the inverse of Eq.~\eqref{tvptvm}. After getting the profile of the velocity, the enthalpy profile is given by Eq.~\eqref{enthalpy} with the boundary condition
\begin{eqnarray}\label{enthalpy boundary}
	\omega_- = \frac{\gamma_w^2 v_w}{\tilde \gamma_-^2 \tilde v_-}\omega_n \,\,.
\end{eqnarray}

For the deflagration mode, we have $v_-=0$, $v_2=0$, and $\alpha_2=\alpha_n$. Or equivalently, $\tilde v_-=\xi_w=v_w$, $\tilde v_2=\xi_{sh}$, and $\alpha_2=\alpha_n$. Eq.~\eqref{velocity continuity} depicts the profile between $\xi_w$ and $\xi_{sh}$, and the boundary condition is $v_+$ in front of the bubble wall or $v_1$ behind the shock front. $v_+=\mu(\xi_w,\tilde v_+)$ with $\tilde v_+$ given by Eq.~\eqref{tvptvm}. However, here $\alpha_+ \neq \alpha_n$ because of the heating effects in front of the bubble wall. In order to solve this, one can use the \emph{shooting method} that first guess the temperature behind the bubble wall $T_-$, then from Eq.~\eqref{vpm1} we get initial values and integrate Eqs.~\eqref{velocity continuity} and \eqref{enthalpy} up to the point where $\mu(v(\xi),\xi)\xi=1/3$ to find the position of the shock front. Then adjust $T_-$ until Eq.~\eqref{shock} is satisfied.

For the hybrid mode, it has both the shock front and the rarefaction wave. We can use the similar procedure as deflagration, except the boundary condition that $\tilde v_-=c_s$. 

In Fig.~\ref{vTvw}, we show the velocity and temperature just in front of and behind the bubble wall and we choose $\alpha_n=0.1$. For deflagrations we have $\tilde v_-=v_w$ and $\tilde v_+<\tilde v_-$. For the hybrid case we have $\tilde v_- = c_s$. For detonations $\tilde v_+=v_w$ and $\tilde v_-<\tilde v_+$. For deflagration and hybrid modes, they are both heating or cooling in front of and behind the bubble wall. However for detonations we have $T_+=T_n$ and only heating behind the bubble wall. As we will see in Sec.~\ref{profile}, $\tilde v_+$ and $T_+$ or $\tilde v_-$ and $T_-$ will give the boundary conditions for velocity and temperature profiles across the bubble wall when the wall has nonzero width.

\begin{figure}[htbp]
	\centering
	\subfigure{
		\begin{minipage}[t]{0.5\linewidth}
			\centering
			\includegraphics[scale=0.45]{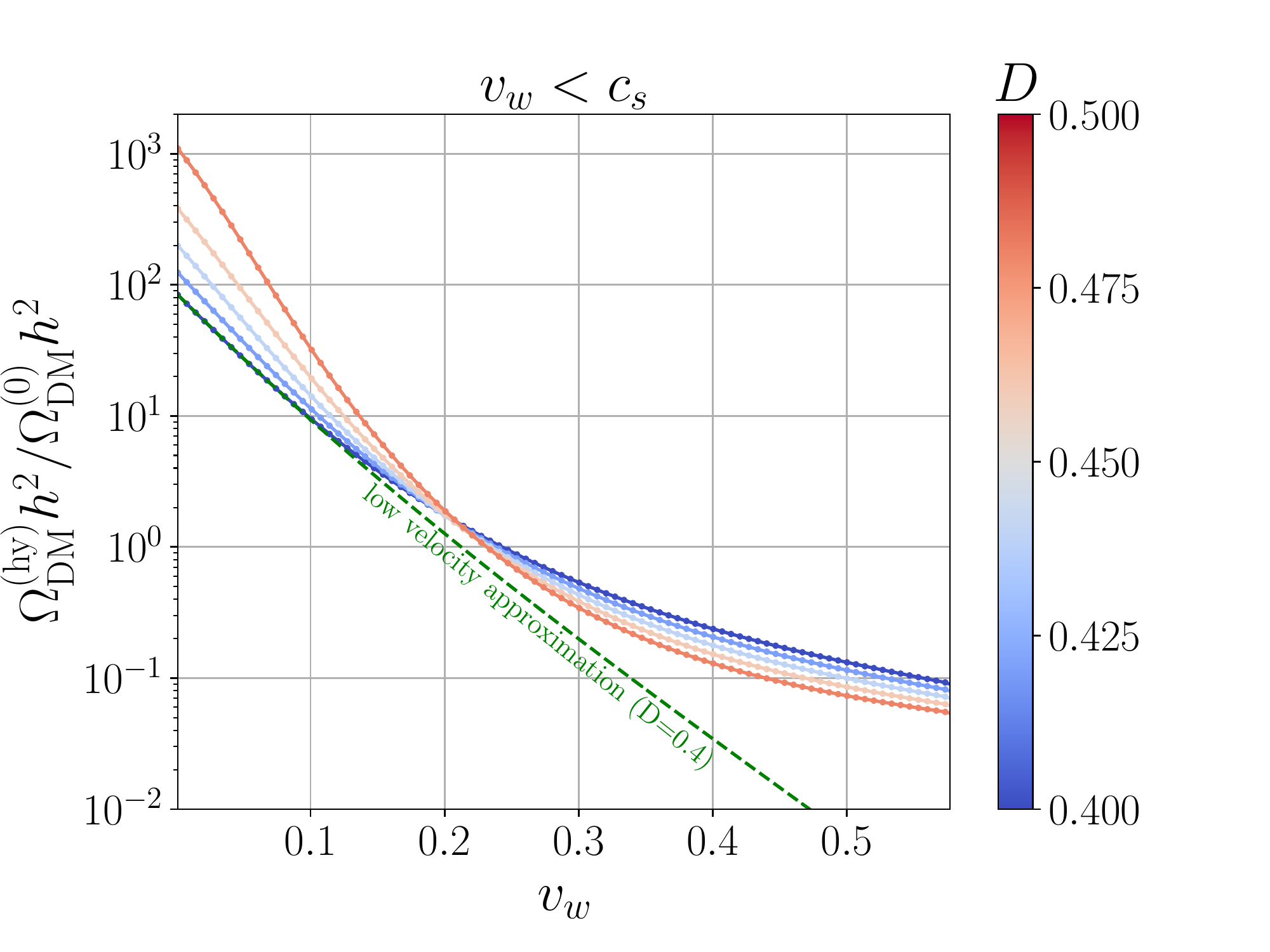}
	\end{minipage}}%
	\subfigure{
		\begin{minipage}[t]{0.55\linewidth}
			\centering
			\includegraphics[scale=0.45]{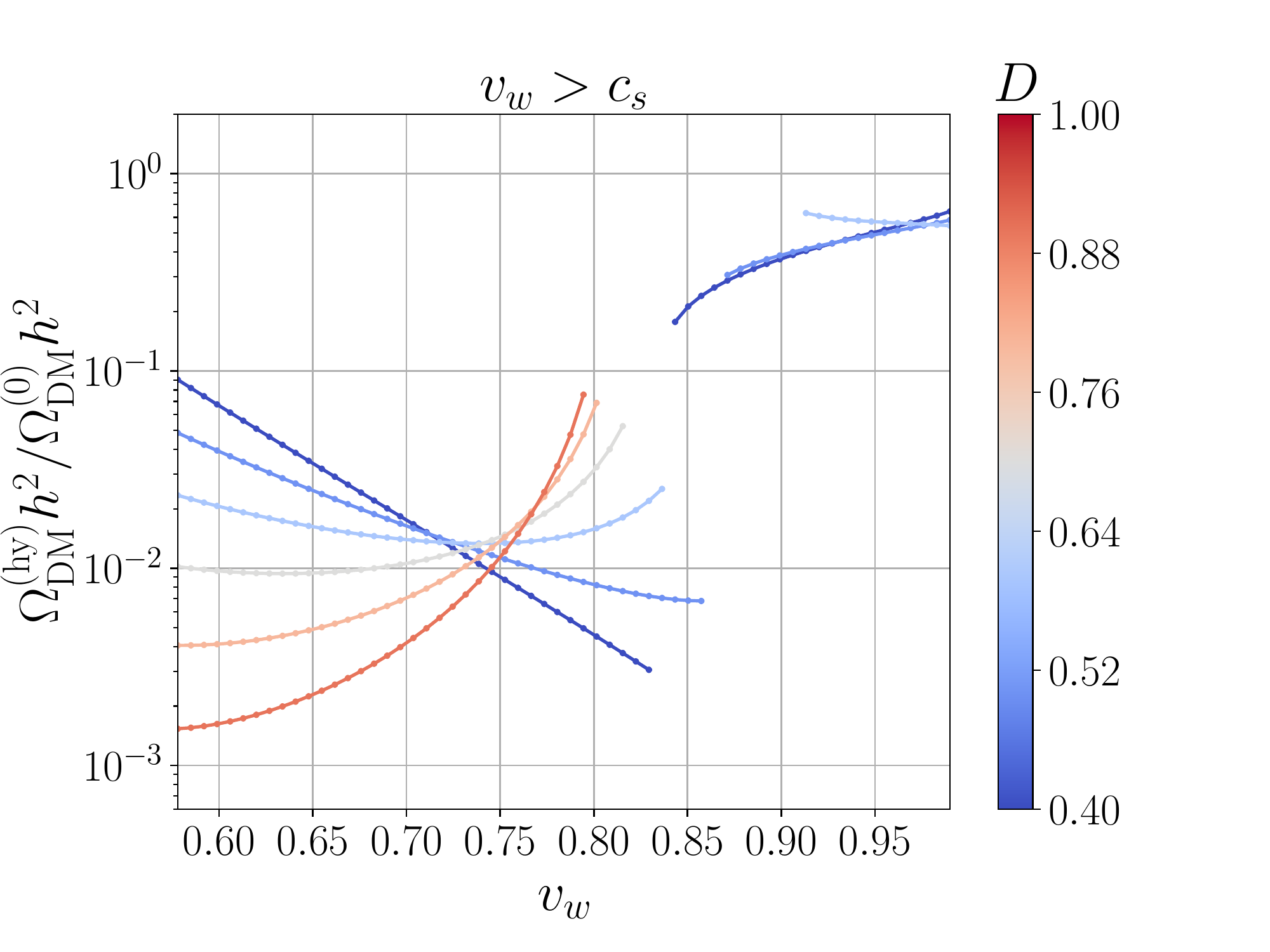}
	\end{minipage}}	
	\caption{Ratio of DM relic density $\Omega_{\mathrm{DM}}^{(\mathrm{hy})} h^2/\Omega_{\mathrm{DM}}^{(0)} h^2$ as a function of $v_w$ and $D$.
		Left is for $v_w<c_s$ (deflagrations) and right is for $v_w>c_s$ (hybrids and detonations).  We fix $C=0.04$ and $\lambda = 0.01$. And we have fixed $y_\chi =3$, which gives $\Omega_{\mathrm{DM}}^{(0)} h^2\simeq 0.12$ at $v_w=0.01$ for $BP_1$.}\label{rinscan}.
\end{figure}

The behavior of $\Omega_{\mathrm{DM}}^{(\mathrm{hy})} h^2/\Omega_{\mathrm{DM}}^{(0)} h^2$, which is the ratio between the DM relic abundance in the three hydrodynamic mode and the DM relic density without heating effects is shown as functions of $v_w$ and $D$ in Fig.~\ref{rinscan}. In the left panel we show the deflagration case for $v_w<c_s$. The effect of increasing $D$ is mainly increasing $\alpha_n$. We can see that for low velocity the DM relic density is enhanced, which is mainly due to the heating effects from which $T_-<T_n<T_+$ such that $\phi_-<\phi_n$. Then the DM is more preferable to pass through the bubble wall. As the velocity increases the effect is weaker. For $v_w \gtrsim 0.1$ we have $\Omega_{\mathrm{DM}}^{(\mathrm{hy})} h^2/\Omega_{\mathrm{DM}}^{(0)} h^2<1$, which is mainly due to the suppression of velocity in front of the bubble wall. This can be seen from $\omega_+\tilde \gamma_+^2 \tilde v_+=\omega_-\tilde \gamma_-^2 \tilde v_-$ that the variation of velocity is opposite to the temperature. And we can see from Eq.~\eqref{nana} that the variation of the velocity will compensate the variation of the temperature. For the detonation case the heating effects influence slightly because the main difference is the vacuum value $\phi_->\phi_n$ such that $m_\chi^{\mathrm{in}}(T_-) >m_\chi^{\mathrm{in}}(T_n) $. The discontinuity in right panel is due to the jump of temperature and velocity that can be seen in Fig.~\ref{vTvw}. And we find for $D>0.7$ that there are no solutions for detonations because the heating is strong enough that $T_->T_c$ and $\phi(T_-)$ is complex.\footnote{This behavior may provide a possibility to achieve super-sonic baryogenesis~\cite{Caprini:2011uz}.} Actually, this is not the case in reality because the heating is strong enough to produce hydrodynamic back-reaction, which gives a upper limit of the bubble wall velocity~\cite{Konstandin:2010dm},
\begin{equation}
	v_w \simeq\left(\frac{\log \frac{T_c}{T_n}}{6 \alpha_c}\right)^{1 / 2} = \left(\frac{T_c^4\log \frac{T_c}{T_n}}{6 T_n^4 \alpha_n}\right)^{1 / 2}\,\,,
\end{equation}
where $\alpha_c$ and $\alpha_n$ are the phase transition strength defined in Eq.~\eqref{PTstrength} with $T_c$ and $T_n$, respectively.

From Eq.~\eqref{tvptvm}, in the small limit of bubble wall velocity we have the approximation
\begin{eqnarray}
	\tilde v_+ \simeq \frac{\tilde v_-(1-2\alpha_+-3\alpha_+^2)}{1+\alpha_+}\,\,.
\end{eqnarray}
Then from Eq.~\eqref{flow} we have
\begin{eqnarray}
	\frac{\omega_-}{\omega_+} = \frac{\tilde \gamma_+^2 \tilde v_+}{\tilde \gamma_-^2 \tilde v_-} \simeq \frac{1-2\alpha_+-3\alpha_+^2}{1+\alpha_+} \equiv \frac{a_-T_-^4}{a_+T_+^4}\,\,.
\end{eqnarray}	
From Fig.~\ref{vTvw} we can see that for $v_w \rightarrow 0$ we have $\alpha_+\approx \alpha_n$ then
\begin{eqnarray}
	\tilde v_+ \simeq \frac{v_w(1-2\alpha_n-3\alpha_n^2)}{1+\alpha_n}, \quad T_- \simeq \left(\frac{1-2\alpha_n-3\alpha_n^2}{1+\alpha_n}\right)^{\frac{1}{4}}\left(\frac{a_+}{a_-}\right)^{\frac{1}{4}} T_n\,\,.
\end{eqnarray}
For example, in the limit that $v_w=0.01$, for $BP_1$ we have $T_-\approx 33~\mathrm{GeV}$ and then from Eq.~\eqref{vevT} we have $\phi(T_-)=438~\mathrm{GeV}$, we can get that $\Omega_{\mathrm{DM}}^{(\mathrm{hy})} h^2/\Omega_{\mathrm{DM}}^{(0)} h^2\simeq 65$, which is consistent with a more exact result without using the low bubble wall velocity limit. This can be seen in Fig.~\ref{rinscan}, where we show the $\Omega_{\mathrm{DM}}^{(\mathrm{hy})} h^2/\Omega_{\mathrm{DM}}^{(0)} h^2$ for different values of $D$ and $v_w$. We fix $C=0.04$ and $\lambda =0.01$. The low velocity approximation for $D=0.4$ is represented by the green dashed line. It is shown that the low velocity approximation fits well with the exact analytic results.

In summary, in the low velocity limit as $v_w\lesssim 0.1$ the quantities around the bubble wall can be evaluated approximately,
\begin{equation}\label{lva}
	\tilde v_-=v_w, \quad \tilde v_+ \simeq \frac{v_w(1-2\alpha_n-3\alpha_n^2)}{1+\alpha_n}, \quad T_+ \simeq T_n, \quad T_- \simeq \left(\frac{1-2\alpha_n-3\alpha_n^2}{1+\alpha_n}\right)^{\frac{1}{4}}\left(\frac{a_+}{a_-}\right)^{\frac{1}{4}} T_n\,\,,
\end{equation}
then the analytic expression of DM number density is
\begin{eqnarray}\label{nana}
	n_\chi^{\text {in }} \simeq \frac{g_{\chi} T_{+}^3}{\gamma_w v_w}\left(\frac{\tilde \gamma_{+}\left(1- \tilde v_{+}\right) m_\chi^{\mathrm{in}} / T_{+}+1}{4 \pi^2 \tilde \gamma_{+}^3\left(1- \tilde v_{+}\right)^2}\right) e^{-\tilde \gamma_{+}\left(1- \tilde v_{+}\right) m_\chi^{\mathrm{in}}\left(T_-\right)/T_{+}} 
\end{eqnarray}
which can be used to get relatively more precise analytic results.

\subsection{Planar approximation}
The hydrodynamic equations can be solved analytically for planar walls. As $j=0$ in Eq.~\eqref{velocity continuity}, we have either $v'(\xi)\equiv 0$ or $\mu(\xi,v)=\pm c_s$. At the bubble wall, the fluid velocity should fulfill $v_{\pm}<\xi_w$ such that  $\mu(\xi,v)=- c_s$ is not physical. Then the physical solutions are either the constants or the rarefaction,
\begin{eqnarray}\label{rarv}
	v_{rar}(\xi)=\frac{\xi - c_s}{1-c_s\xi}\,\,.
\end{eqnarray} 
The integration in Eq.~\eqref{enthalpy} is simple to do. For $v'(\xi) =0$ the enthalpy is a constant. For $\mu(\xi,v)=c_s$ we have 
\begin{eqnarray}\label{planar enthalpy}
	\frac{\omega}{\omega_0} = \left(\frac{1-v_0}{1+v_0}\frac{1+v}{1-v}\right)^{2/\sqrt{3}}\,\,.
\end{eqnarray}

For detonations, the fluid velocity between $\xi=c_s$ and a certain $\xi_0 \leq \xi_w$ is given by $v(\xi)=v_{rar}(\xi)$. And the fluid velocity between $\xi_0$ and $\xi_w$ is constant $v\equiv v_-$. From the matching condition $v_{rar}(\xi_0)=v_-$ we can determine the position $\xi_0$,
\begin{eqnarray}
	\xi_0=\frac{v_-+c_s}{1+v_-c_s}\,\,.
\end{eqnarray}
$v_-=\mu(\xi_w,\tilde v_-)$ with $\tilde v_-$ is given by  the inverse of Eq.~\eqref{tvptvm} as a function of $\alpha_+=\alpha_n$ and $\tilde v_+=\xi_w$. Between $c_s$ and $\xi_0$, the enthalpy profile is given by Eq.~\eqref{planar enthalpy} with the boundary condition $\omega_{0}=\omega_-$. The $\omega_-$ is given by Eq.~\eqref{enthalpy boundary} and is also the constant value between $\xi_0$ and $\xi_w$. Inserting Eq.~\eqref{rarv} into Eq.~\eqref{planar enthalpy}, we have
\begin{eqnarray}
	\omega(\xi) = \omega_- \left(\frac{1-v_-}{1+v_-}\frac{1-c_s}{1+c_s}\frac{1+\xi}{1-\xi}\right)^{2/\sqrt{3}}\,\,.
\end{eqnarray}

It is much simpler for deflagration mode. In this case, the fluid velocity is constant $v\equiv v_+=v_1$ between $\xi_w$, and $\xi_{sh}$ and vanishes outside this region. We also have $\tilde v_-=\xi_w$, $\tilde v_2=\xi_{sh}$, $\omega_+=\omega_1$ and $\alpha_+=\alpha_1$. Then by using Eq.~\eqref{shock}, we have
\begin{eqnarray}\label{defwv}
	3\xi_{sh}^2=\frac{3\alpha_n+\alpha_1}{\alpha_n+3\alpha_1},\quad v_+=v_1=\frac{3\xi_{sh}^2-1}{2\xi_{sh}}\,\,.
\end{eqnarray}
Combine this with Eq.~\eqref{tvptvm}
we can get $\xi_{sh}$ as a function of $\alpha_n$ and $\xi_w \equiv v_w$,
\begin{eqnarray}\label{nshock}
	(3\xi_{sh}^2-1)^2+\xi_{sh}(3\xi_{sh}^2-1)\frac{1-3\xi_w^2}{\xi_w}=\frac{9}{2}\alpha_n (1-\xi_{sh}^2)^2
\end{eqnarray}
which we can solve numerically.

For the hybrid case we get a similar equation as Eq.~\eqref{nshock} by using the condition $v_+=v_1$ and $\alpha_+=\alpha_1$,
\begin{eqnarray}\label{nshock2}
	\left[\xi_{sh}(1-\sqrt{3}\xi_w)-\frac{3\xi_{sh}^2-1}{2}(\xi_w-\sqrt{3})\right]^2=\frac{9}{4}\alpha_n (1-\xi_w^2)(1-\xi_{sh}^2)^2\,\,.
\end{eqnarray}

\begin{figure}[htbp]
	\begin{center}
		\includegraphics[scale=0.42]{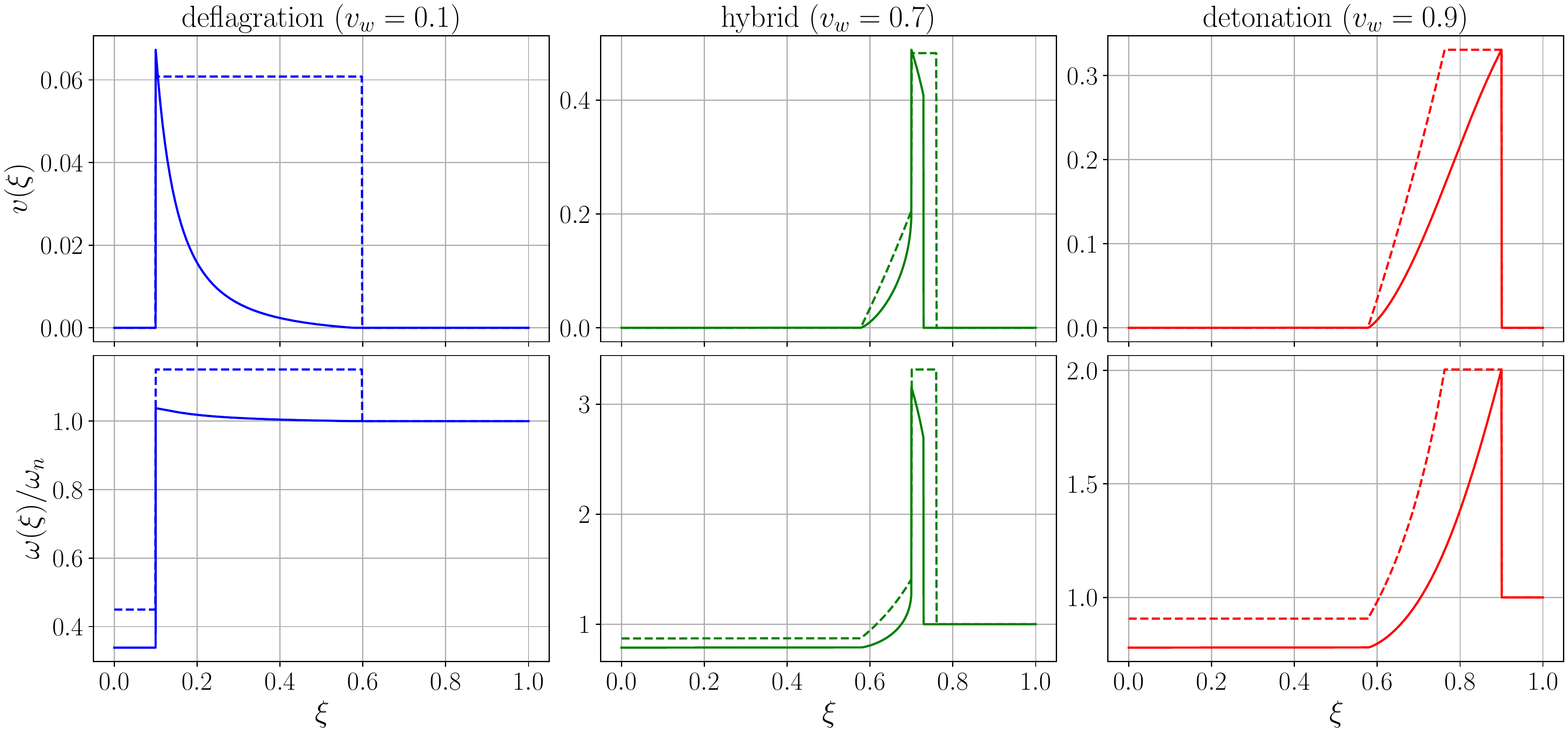}
		\caption{Velocity and enthalpy profiles of three hydrodynamic modes for $\alpha_n =0.23$. The solid lines are for spherical bubbles and dashed lines are for the approximation of the planar wall. }
		\label{vw}
	\end{center}
\end{figure}

In Fig.~\ref{vw}, we show the velocity and enthalpy profile in both spherical and planar cases for $\alpha_n =0.23$.
The solid lines represents the results for spherical bubbles and the 
 dashed lines are for the approximation of planar wall. 

\section{Temperature and velocity profile with bubble wall  thickness}\label{Tvprofile}
Further, we relax the assumption of zero-width bubble wall and consider the wall thickness, which 
can help to perform more precise results with Boltzmann equations in next section.
As the bubble wall has a finite width, we expect the velocities and the temperature of the plasma have a profile across the wall. Below we neglect the index $w$ of $z^w$ and $p_z^w$ for simplicity. Keep in mind that we work in the bubble wall frame.

To derive the equations that describe the evolution of $T(z)$ and $\tilde v_{\mathrm{pl}}(z)$, we can still use the equations of conservation of EMT $\nabla_\mu T^{\mu \nu}=0$~\cite{Laurent:2022jrs}
\begin{eqnarray}\label{EMT}
\begin{aligned}
	T^{30} & =\omega_{\mathrm{pl}} \tilde \gamma_{\mathrm{pl}}^2 \tilde v_{\mathrm{pl}}\equiv c_1 \,\,,\\
	T^{33} & =\frac{1}{2}\left(\partial_{z} \phi\right)^2-V_{\rm eff}\left(\phi, T\right)+\omega_{\mathrm{pl}}  \tilde\gamma_{\mathrm{pl}}^2 \tilde v_{\mathrm{pl}}^2 \equiv c_2 \,\,,
\end{aligned}
\end{eqnarray}
where $c_1$ and $c_2$ are constants that depend on boundary values $T_{-}$ and $\tilde v_{-}$ (or alternatively on $T_{+}$ and $\tilde v_{+}$), which denote the fluid temperature and velocity at $z\rightarrow - \infty$ ($+\infty$). It is simple to show that, if we parametrize the $V_{\rm eff}$ as 
\begin{eqnarray}
	V_{\rm eff} = -p = \epsilon - aT^4/3\,\,,
\end{eqnarray}
this is just the bag model case. However, in reality the $a$ and $\epsilon$ is functions of temperature which will give us results beyond bag model~\cite{Leitao:2014pda,Giese:2020rtr,Wang:2020nzm}.

In practice, one can directly solve the first line of Eq.~\eqref{EMT} for $v_{\mathrm{pl}}$ :
\begin{eqnarray}\label{vpro}
\tilde v_{\mathrm {pl}}=\frac{-\omega_{\mathrm{pl}}+\sqrt{4 c_1^2+\omega_{\mathrm{pl}}^2}}{2 c_1}\,\,.
\end{eqnarray}
Substituting $\tilde v_{\mathrm{pl}}$ into the equation for $T^{33}$ yields
\begin{eqnarray}\label{Tpro}
\frac{1}{2}\left(\partial_{z} \phi\right)^2-V_{\rm eff}-\frac{1}{2} \omega_{\mathrm{pl}}+\frac{1}{2} \sqrt{4 c_1^2+\omega_{\mathrm{pl}}^2}-c_2=0\,\,.
\end{eqnarray}
In order to get $c_1$ and $c_2$, we should solve for hydrodynamic equations for the bubble wall system just as in last section, in particular, for three cases: detonations, deflagrations, and hybrids. As we can see in Fig.~\ref{vw}, we can in many cases just use the planar approximation. Then for three cases, we have

1) Detonations: $T_+ = T_n$ and $\tilde v_+=v_w$, which gives us 
\begin{eqnarray}
	c_1 = \omega_n \gamma_w^2 v_w, \quad c_2 = \omega_n \gamma_w^2 v_w^2 - V_{\rm eff}(0,T_n)\,\,.
\end{eqnarray}

2) Deflagrations: After getting $\xi_{ sh}$ from Eq.~\eqref{nshock}, we have from Eq.~\eqref{defwv}
\begin{eqnarray}
	\omega_+ = \omega_n \frac{9\xi_{sh}^2}{3(1-\xi_{sh}^2)}, \quad \tilde v_+ = \frac{v_w - v_+}{1-v_w v_+} \quad \text{where} \quad v_{+}=\frac{3\xi_{sh}^2-1}{2\xi_{sh}}\,\,,
\end{eqnarray}
then
\begin{eqnarray}
	c_1 = \omega_+\tilde \gamma_+^2 \tilde v_+, \quad c_2 = \omega_+ \tilde \gamma_+^2 \tilde v_+^2 - V_{\rm eff}(0,T_+)\,\,.
\end{eqnarray}

3) Hybrids: After solving $\xi_{sh}$ from Eq.~\eqref{nshock2}, we have the same form of $c_1$ and $c_2$ as deflagrations.

After getting $c_1$ and $c_2$, we can solve Eqs.~\eqref{vpro} and \eqref{Tpro} to get the profiles of velocity $\tilde v_{\mathrm{pl}}(z)$ and temperature $T(z)$ around the bubble wall.

\begin{figure}[htbp]
	\centering
	\subfigure{
		\begin{minipage}[t]{0.5\linewidth}
			\centering
			\includegraphics[scale=0.4]{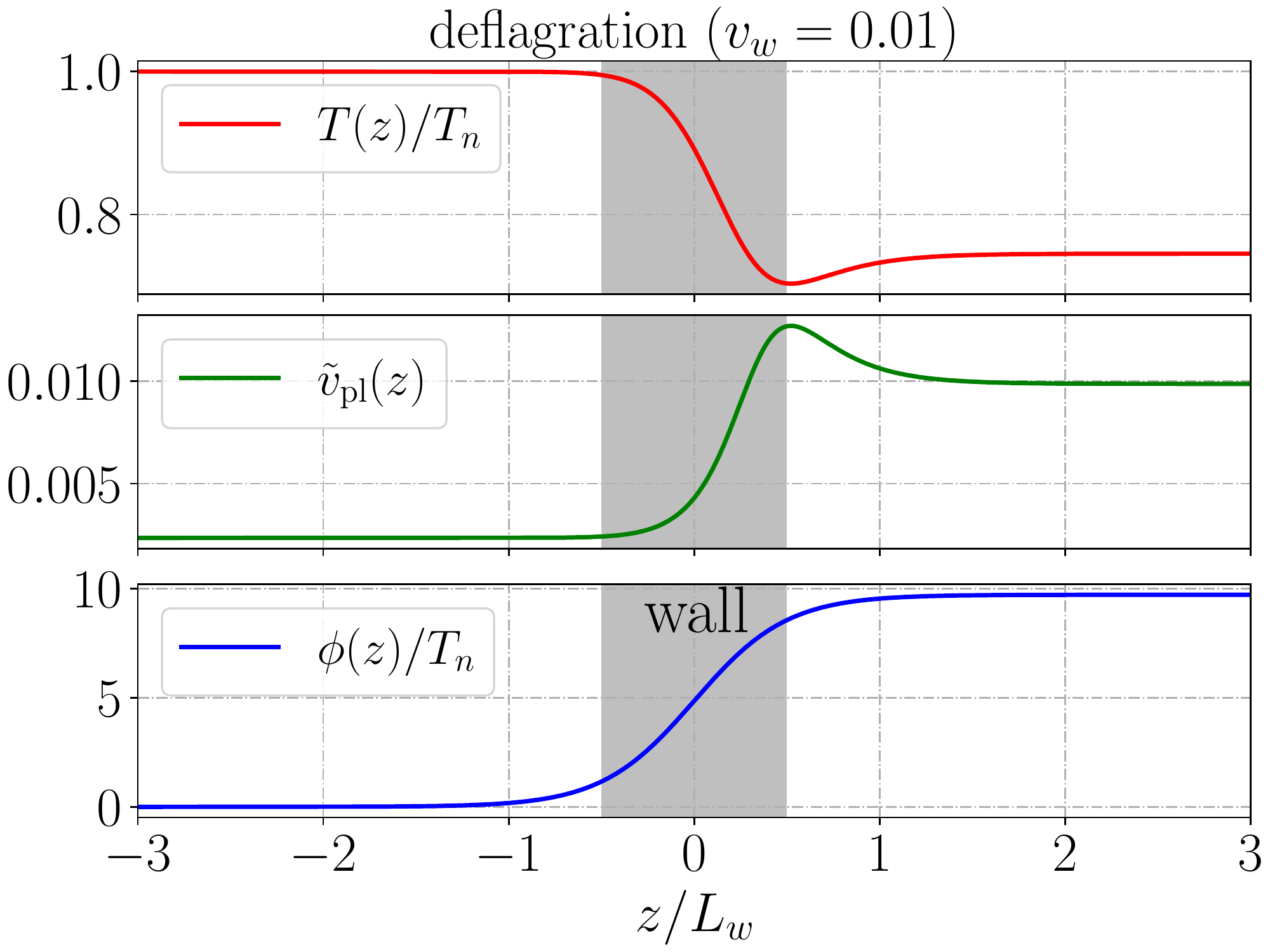}
	\end{minipage}}%
	\subfigure{
		\begin{minipage}[t]{0.5\linewidth}
			\centering
			\includegraphics[scale=0.4]{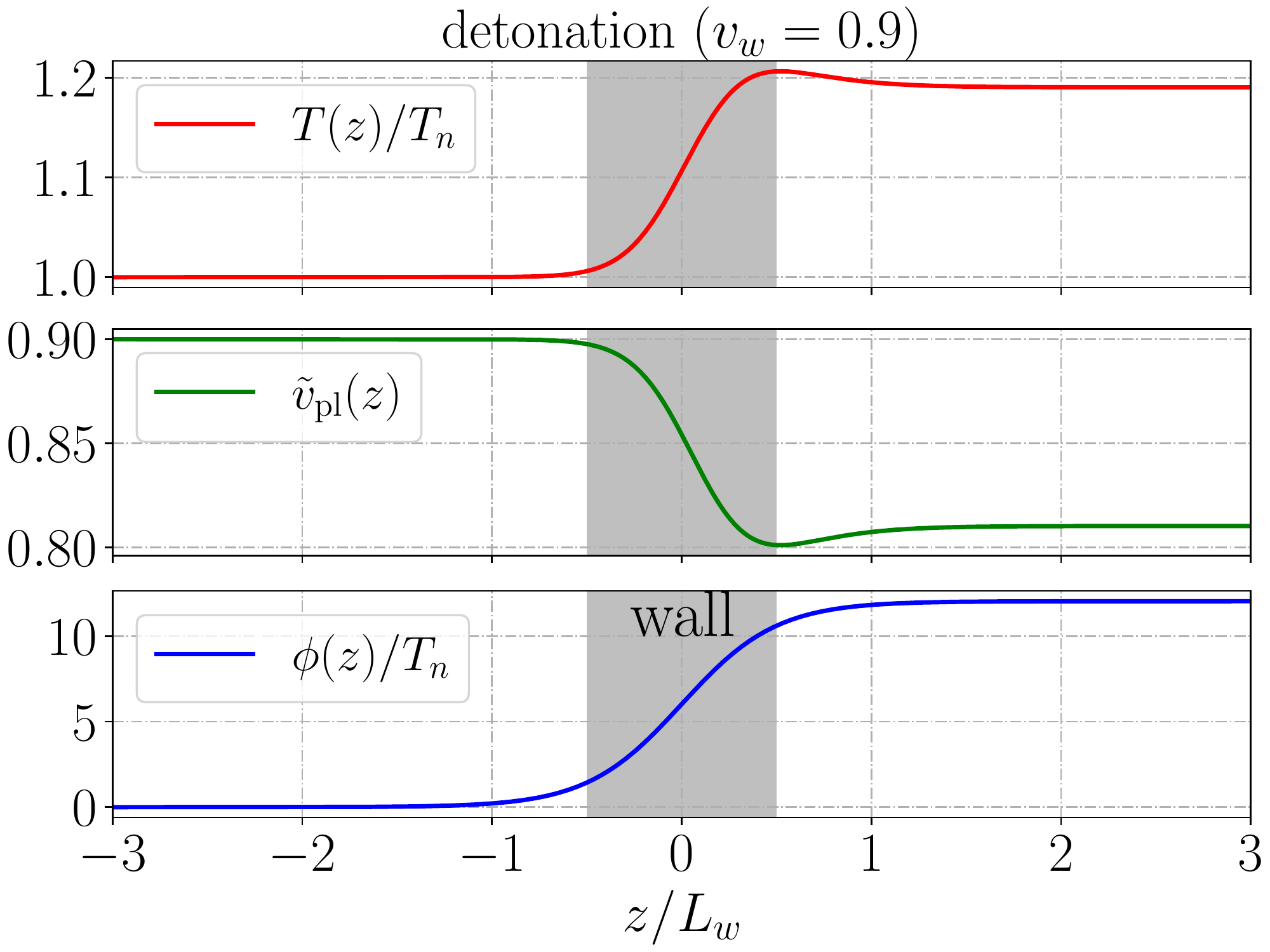}
	\end{minipage}}	
	\caption{Velocity and temperature profiles of DM across the bubble wall for deflagration ($v_w=0.01$ in the left panel) and detonation ($v_w=0.9$ in the right panel).
	The hydrodynamic effects and bubble width are considered.}\label{profile}.
\end{figure}

In addition to Eqs.~\eqref{vpro} and \eqref{Tpro}, we have another equation, i.e. the EOM of the background scalar field,
\begin{eqnarray}
	\partial^2\phi +\frac{\partial V_{\rm eff}(\phi,T)}{\partial \phi}=0\,\,.
\end{eqnarray}
A simple solution of this equation is \emph{tanh} ansatz,
\begin{eqnarray}
	\phi(z)=\frac{\phi(T_-)}{2}\left(1+{\rm tanh} \frac{2z}{L_w}\right) \,\,.
\end{eqnarray}
We choose the wall width $L_w=5/T_n$.
We show the results of $T(z)$ and $\tilde v_{\mathrm{pl}}(z)$ in Fig.~\ref{profile}, where we choose the relatively low velocity $v_w=0.01$ and the relatively high velocity $v_w=0.9$, respectively. The left figure represents the deflagration mode where the temperature inside the bubble is lower than $T_n$. We can also see that $\tilde v_- = v_w$ which fulfills the boundary condition for deflagration.
In contrast, the right figure depicts the detonation mode where the temperature in front of the bubble wall $T_+=T_n$ and temperature inside bubble is higher than $T_n$. For detonation we also have $\tilde v_+=v_w$.

\section{Numerical results of the hydrodynamic effects on Boltzmann equation with bubble wall thickness}\label{boltzmann}
Generally the bubble wall is not extremely thin. The distribution of particles varies across the bubble wall caused by the gradient of their masses. To obtain more precise results of the hydrodynamic effects on the DM density, we perform numerical calculations with the Boltzmann equation and bubble wall thickness.
\subsection{Boltzmann equation}
The Boltzmann equation characterizes the evolution of particles in phase space,
\begin{eqnarray}
\mathbf{L}\left[f_\chi\right]=\mathbf{C}\left[f_\chi\right]\,\,.
\end{eqnarray}
Here, $\mathbf{L}$ is the Liouville operator and $\mathbf{C}$ is a collision term, which accounts for the particle interactions.

We introduce the fluid ansatz, the distribution function of DM in the wall frame is parametrized as
\begin{eqnarray}
	f_\chi=\mathcal{A}\left(z, p_z\right) f_{\chi,+}^{\rm eq}=\mathcal{A}\left(z, p_z\right) \exp \left(-\frac{\tilde \gamma_{+}(E-\tilde v_{+} p_z)}{T_{+}}\right)\,\,,
\end{eqnarray}
where $ f_{\chi,+}^{\rm eq}$ is the equilibrium distribution function in front of the bubble wall. Because of the energy conservation, in the bubble wall frame $E=\sqrt{p_x^2+p_y^2+p_z(z)^2+m_\chi(z)^2}\equiv \text{const.}$ with
\begin{eqnarray}
	m_\chi(z) \equiv \frac{	m_\chi^{\mathrm{in}}(\phi_-)}{2}\left(1+\mathrm{tanh} \frac{2z}{L_w}\right)\,\,,
\end{eqnarray}
where $m_\chi^{\mathrm{in}}(\phi_-)=y_\chi \phi_-$ is the mass of DM deep inside the bubble.

The Liouville operator is the total time derivative of the phase space distribution function $f=f(t, \mathbf{x}(t), \mathbf{p}(t))$. In a steady state and neglect the $x$, $y$ direction, using energy-conservation $p_z^2+m_\chi(z)^2 = \mathrm{const.}$ in the wall frame, 
we then have
\begin{eqnarray}
	\mathbf{L}\left[f_\chi\right]=\frac{p_z}{E} \frac{\partial f_\chi}{\partial z}-\frac{m_\chi}{E} \frac{\partial m_\chi}{\partial z} \frac{\partial f_\chi}{\partial p_z} \,\,.
\end{eqnarray}
Then we find that the nonzero $\partial m_\chi/\partial z$ acts as a external force which disturb the phase space distribution of DM.
We now integrate over the transverse momentum components $p_x$ and $p_y$ and multiply by the number of spin states, $g_\chi=2$, giving
\begin{eqnarray}
	g_\chi \int \frac{d p_x d p_y}{(2 \pi)^2} \mathbf{L}\left[f_\chi\right]=g_\chi \int \frac{d p_x d p_y}{(2 \pi)^2} \frac{p_z}{E} \frac{\partial f_\chi}{\partial z}-g_\chi\left(\frac{\partial m_\chi}{\partial z}\right) \int \frac{d p_x d p_y}{(2 \pi)^2} \frac{m_\chi}{E} \frac{\partial f_\chi}{\partial p_z} \,\,.
\end{eqnarray} 
Using the Maxwell-Boltzmann approximation for $f_\chi$ then gives
\begin{eqnarray}
	\begin{aligned}
		g_\chi \int \frac{d p_x d p_y}{(2 \pi)^2} \mathbf{L}\left[f_\chi\right] \approx & g_\chi\left(\frac{\partial}{\partial z} \mathcal{A}\left(z, p_z\right)\right) \int \frac{d p_x d p_y}{(2 \pi)^2} \frac{p_z}{\sqrt{m_\chi^2+p_x^2+p_y^2+p_z^2}} e^{-E_+^\mathcal{P} / T_{+}} \\
		& +g_\chi \mathcal{A}\left(z, p_z\right) \int \frac{d p_x d p_y}{(2 \pi)^2} \frac{p_z}{\sqrt{m_\chi^2+p_x^2+p_y^2+p_z^2}}\left(\frac{\partial}{\partial z} e^{-E_+^\mathcal{P} / T_{+}}\right) \\
		& -g_\chi\left(\frac{\partial m_\chi}{\partial z}\right)\left(\frac{\partial}{\partial p_z} \mathcal{A}\left(z, p_z\right)\right) \int \frac{d p_x d p_y}{(2 \pi)^2} \frac{m_\chi}{\sqrt{m_\chi^2+p_x^2+p_y^2+p_z^2}} e^{-E_+^\mathcal{P} / T_{+}} \\
		& -g_\chi\left(\frac{\partial m_\chi}{\partial z}\right) \mathcal{A}\left(z, p_z\right) \int \frac{d p_x d p_y}{(2 \pi)^2} \frac{m_\chi}{\sqrt{m_\chi^2+p_x^2+p_y^2+p_z^2}}\left(\frac{\partial}{\partial p_z} e^{-E_+^\mathcal{P} / T_{+}}\right)\,\,, 
	\end{aligned}
\end{eqnarray}
where
\begin{eqnarray}
	E^\mathcal{P}_+ \equiv E^\mathcal{P}_+(z,p_z)= \tilde\gamma_{+}(E-\tilde v_{+} p_z)\,\,.
\end{eqnarray}

After the integration the Liouville term becomes
\begin{eqnarray}\label{liou}
	\begin{aligned}
		&g_\chi \int \frac{d p_x d p_y}{(2 \pi)^2} \mathbf{L}\left[f_\chi\right] \approx \\
		&\left[\left(\frac{p_z}{m_\chi} \frac{\partial}{\partial z}-\left(\frac{\partial m_\chi}{\partial z}\right) \frac{\partial}{\partial p_z}-\left(\frac{\partial m_\chi}{\partial z}\right) \frac{\tilde \gamma_{+}\tilde v_{+}}{T_{+}}\right) \mathcal{A}\left(z, p_z\right)\right] \frac{g_\chi m_\chi T_{+}}{2 \pi \tilde\gamma_{+}} e^{\tilde\gamma_{+}\left(\tilde v_{+} p_z-\sqrt{m_\chi^2+p_z^2}\right) / T_{+}} \,\,.
	\end{aligned}
\end{eqnarray}

\subsection{Collision term}
One can evaluate the collision term $\mathbf{C}\left[f_\chi\right]$ in the local plasma frame of DM and turn back to wall frame at the end. We consider the process of annihilation $\chi\left(p^\mathcal{P}\right)+\bar{\chi}\left(q^\mathcal{P}\right) \rightarrow \phi\left(k^\mathcal{P}\right)+\phi\left(l^\mathcal{P}\right)$ and  inverse decay $\chi\left(p^\mathcal{P}\right)+\bar{\chi}\left(q^\mathcal{P}\right) \rightarrow \phi\left(k^\mathcal{P}\right)$ in our case.  The collision terms for the other processes, $\chi \phi \rightarrow \chi \phi$, $\chi \chi \rightarrow \chi \chi$, and $\chi \bar\chi \rightarrow \chi \bar\chi$, has similar expressions. For annihilation, integrating over $p_x$ and $p_y$ and multiplying by the number of spin states, $g_\chi=2$, the collision term is
\begin{eqnarray}
\begin{aligned}
	g_\chi \int \frac{d p_x d p_y}{(2 \pi)^2} \mathbf{C}\left[f_\chi\right]=- & \sum_{\text {spins }} \int \frac{d p_x d p_y}{(2 \pi)^2} d \Pi_{q^\mathcal{P}} d \Pi_{k^\mathcal{P}} d \Pi_{l^\mathcal{P}} \frac{(2 \pi)^4}{2 E_p^\mathcal{P}} \delta^{(4)}\left(p^\mathcal{P}+q^\mathcal{P}-k^\mathcal{P}-l^\mathcal{P}\right)|\mathcal{M}|_{\chi \bar{\chi} \leftrightarrow \phi \phi}^2 \\
	& \times\left[f_{\chi_p} f_{\bar{\chi}_q}\left(1 \pm f_{\phi_k}\right)\left(1 \pm f_{\phi_l}\right)-f_{\phi_k} f_{\phi_l}\left(1 \pm f_{\chi_p}\right)\left(1 \pm f_{\bar{\chi}_q}\right)\right]\,\,,
\end{aligned}
\end{eqnarray}
where $\mathcal{M}_{\chi \bar{\chi} \leftrightarrow \phi \phi}$ is the scattering matrix element, and we have used the notation $E_p^\mathcal{P}=\left[\left(\mathbf{p}^\mathcal{P}\right)^2+m_\chi^2\right]^2$, $d \Pi_{q^\mathcal{P}} \equiv$ $d^3 q^\mathcal{P} /\left[2 E_q^\mathcal{P}(2 \pi)^3\right]$, and $f_{\chi_p} \equiv f_\chi\left(t^\mathcal{P}, \mathbf{x}^\mathcal{P}, \mathbf{p}^\mathcal{P}\right)$, with $\chi_p \equiv \chi(p)$.

We neglect Pauli blocking and Bose enhancement for all species by setting $1 \pm f \approx 1$, and assume that all species except for the initial DM particle $\chi(p)$ are in equilibrium. In the local frame of DM the light particles $\phi$  should obey the equilibrium distribution 
\begin{eqnarray}
f_{\phi}^{\rm eq} =	\exp \left(-\frac{\tilde \gamma_{\rm pl}^{\prime}(z)(E_\phi-\tilde v_{\rm pl}^{\prime}(z) p_{z\phi})}{T_{}(z)}\right) \,\,,
\end{eqnarray}
where
\begin{eqnarray}
   \tilde v_{\rm pl}^{\prime}(z) = \frac{\tilde v_{\rm pl}(z)-\tilde v_+}{1-\tilde v_{\rm pl}(z)\tilde v_+}
\end{eqnarray}
represents the relativistic velocity between $\phi$ and DM.

Since detailed balance holds for each momentum mode independently, $f_{\phi_k}^{\mathrm{eq}} f_{\phi_l}^{\mathrm{eq}}=f_{\chi_p}^{\mathrm{eq}} f_{\bar{\chi}_q}^{\mathrm{eq}}$. 
\begin{equation}
\begin{aligned}
g_\chi \int \frac{d p_x d p_y}{(2 \pi)^2} \mathbf{C}\left[f_\chi\right]=-g_\chi \int \frac{d p_x d p_y}{(2 \pi)^2} d \Pi_{q^\mathcal{P}} d \Pi_{k^\mathcal{P}} d \Pi_{l^\mathcal{P}} \frac{(2 \pi)^4}{2 E_p^\mathcal{P}} \delta^{(4)}\left(p^\mathcal{P}+q^\mathcal{P}-k^\mathcal{P}-l^\mathcal{P}\right)|\mathcal{M}|_{\chi \bar{\chi} \leftrightarrow \phi \phi}^2 \\
\times \left[f_{\chi_p} f_{\bar{\chi}_q,+}^{\mathrm{eq}}-f_{\chi_p}^{\mathrm{eq}} f_{\bar{\chi}_q}^{\mathrm{eq}}\right] \,\,.
\end{aligned}
\end{equation}
Here we approximate $f_{\bar{\chi}_q} \simeq f_{\bar{\chi}_q,+}^{\mathrm{eq}} = \exp(-E_p^\mathcal{P}/T_+)$. 
We will see later that the $\mathcal A$ is of $\mathcal O(1)$ except for some finite regions, so this approximation does not influence the results dramatically.
 
We can now integrate over $k$ and $l$ to obtain
\begin{eqnarray}\label{coll}
\begin{aligned}
	g_\chi \int \frac{d p_x d p_y}{(2 \pi)^2} \mathbf{C}\left[f_\chi\right] & =-g_\chi g_{\bar{\chi}} \int \frac{d p_x d p_y}{(2 \pi)^2 2 E_p^\mathcal{P}} d \Pi_{q^\mathcal{P}} 4 F \sigma_{\chi \bar{\chi} \rightarrow \phi \phi}\left[f_{\chi_p} f_{\bar{\chi}_q,+}^{\mathrm{eq}}-f_{\chi_p}^{\mathrm{eq}} f_{\bar{\chi}_q}^{\mathrm{eq}}\right] \\
	& =-g_\chi g_{\bar{\chi}} \int \frac{d p_x d p_y}{(2 \pi)^2 2 E_p^\mathcal{P}} d \Pi_{q^\mathcal{P}} 4 F \sigma_{\chi \bar{\chi} \rightarrow \phi \phi}\left[\mathcal{A}f_{\chi_p,+}^{\mathrm{eq}} f_{\bar{\chi}_q,+}^{\mathrm{eq}}-f_{\chi_p}^{\mathrm{eq}} f_{\bar{\chi}_q}^{\mathrm{eq}}\right] \\
	& \equiv \Gamma_{\rm P}(z,p_z) \mathcal{A}\left(z, p_z\right)-\Gamma_{\rm I}(z,p_z) \,\,,
\end{aligned}
\end{eqnarray}
where the kinematic factor,
\begin{eqnarray}
F \equiv E_p^\mathcal{P} E_q^\mathcal{P}\left|v_\chi-v_{\bar{\chi}}\right|=\frac{1}{2} \sqrt{\left(s-2 m_\chi^2\right)^2-4 m_\chi^4}\,\,,
\end{eqnarray}
and here the collision rate for $\chi \bar{\chi} \rightarrow \phi \phi$ is defined as
\begin{eqnarray}\label{gammar}
\Gamma^{\mathcal{P}}_{\rm P}(z,E_p^{\mathcal{P}}) =	-g_\chi g_{\bar{\chi}} \int \frac{d p_x d p_y}{(2 \pi)^2 2 E_p^\mathcal{P}} d \Pi_{q^\mathcal{P}} 4 F \sigma_{\chi \bar{\chi} \rightarrow \phi \phi}f_{\chi_p,+}^{\mathrm{eq}} f_{\bar{\chi}_q,+}^{\mathrm{eq}} \,\,,
\end{eqnarray}
and collision rate for their inverse process
\begin{eqnarray}\label{gammai}
	\Gamma^{\mathcal{P}}_{\rm I}(z,E_p^{\mathcal{P}}) =	-g_\chi g_{\bar{\chi}} \int \frac{d p_x d p_y}{(2 \pi)^2 2 E_p^\mathcal{P}} d \Pi_{q^\mathcal{P}} 4 F \sigma_{\chi \bar{\chi} \rightarrow \phi \phi}f_{\chi_p}^{\mathrm{eq}} f_{\bar{\chi}_q}^{\mathrm{eq}} \,\,.
\end{eqnarray}
The integration can be done numerically by using the Monte Carlo package VEGAS~\cite{Hahn:2004fe}. To boost back into the bubble wall frame, in Eqs.~\eqref{gammar} and \eqref{gammai} we should replace $E^\mathcal{P}_p$ with $\tilde \gamma_{+}\left(E-\tilde v_{+} p_z \right)$,
\begin{equation}
\begin{aligned}	
	\Gamma_{\rm P(I)}(z,p_z) = \Gamma^{\mathcal{P}}_{\rm P(I)}\left(z,\tilde \gamma_{+}\left(E-\tilde v_{+} p_z \right)\right)\,\,.
\end{aligned}	
\end{equation}

In the case without heating, we have $\tilde v_+ =v_w$ and $T_+=T_n$, then one can get 
\begin{eqnarray}
	g_\chi \int \frac{d p_x d p_y}{(2 \pi)^2} \mathbf{C}\left[f_\chi\right] = \Gamma_n(z,p_z)(\mathcal A(z,p_z)-1)
\end{eqnarray}
with $\Gamma_{\rm P} = \Gamma_{\rm I} = \Gamma_n$. This is the original form of Ref.~\cite{Baker:2019ndr}.

Outside the bubble, where $m_\chi \approx 0$, the annihilation cross section is
\begin{eqnarray}
\sigma(\chi \bar{\chi} \rightarrow \phi \phi)=\frac{y_\chi^4}{32 \pi s}\left[2 \log \left(s / m_\phi^2\right)-3\right]+\mathcal{O}\left(m_\phi^2 / s\right)\,\,.
\end{eqnarray}

\begin{figure}[htbp]
	\centering
	\subfigure{
		\begin{minipage}[t]{0.5\linewidth}
			\centering
			\includegraphics[scale=0.43]{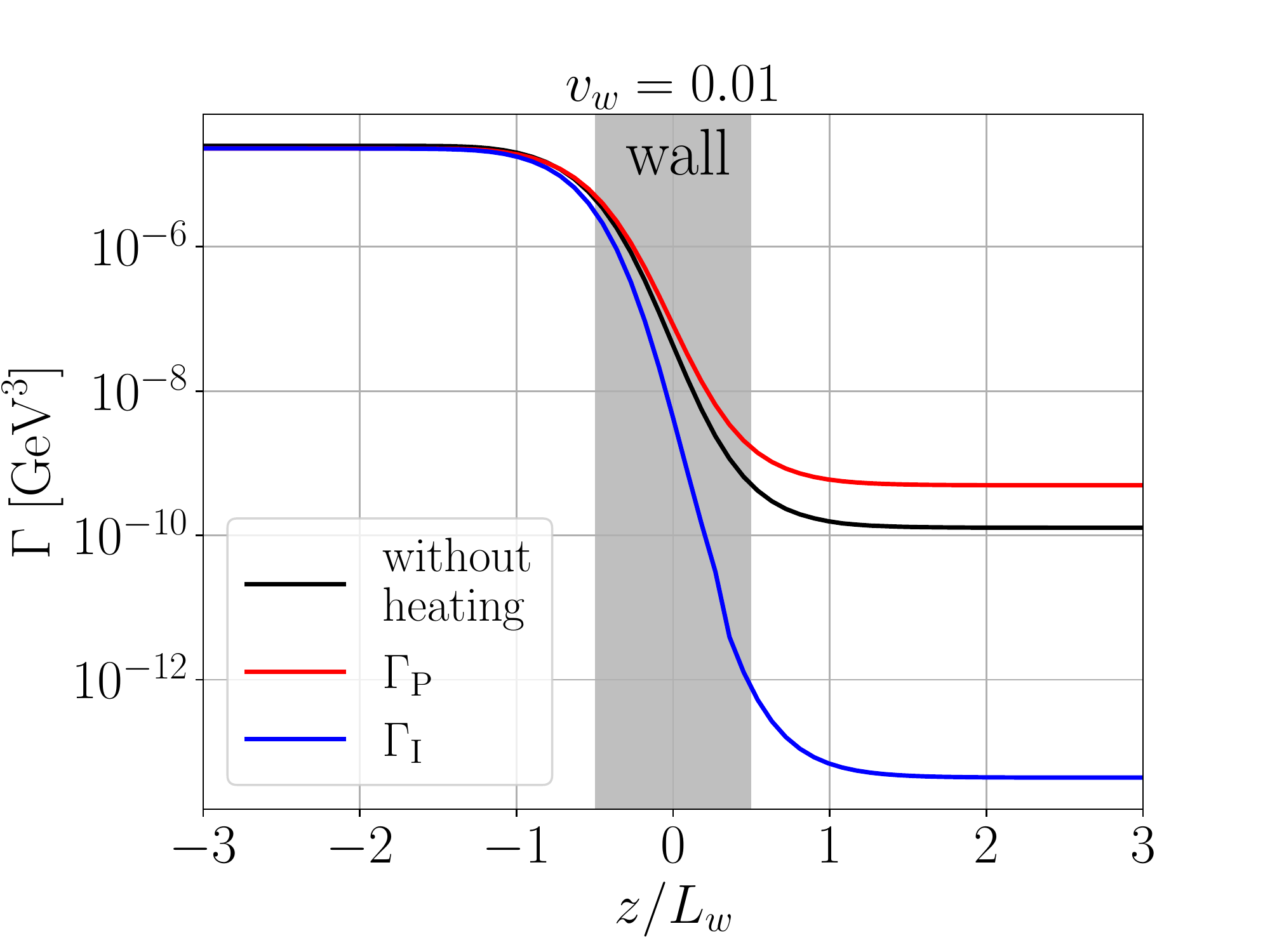}
	\end{minipage}}%
	\subfigure{
		\begin{minipage}[t]{0.5\linewidth}
			\centering
			\includegraphics[scale=0.43]{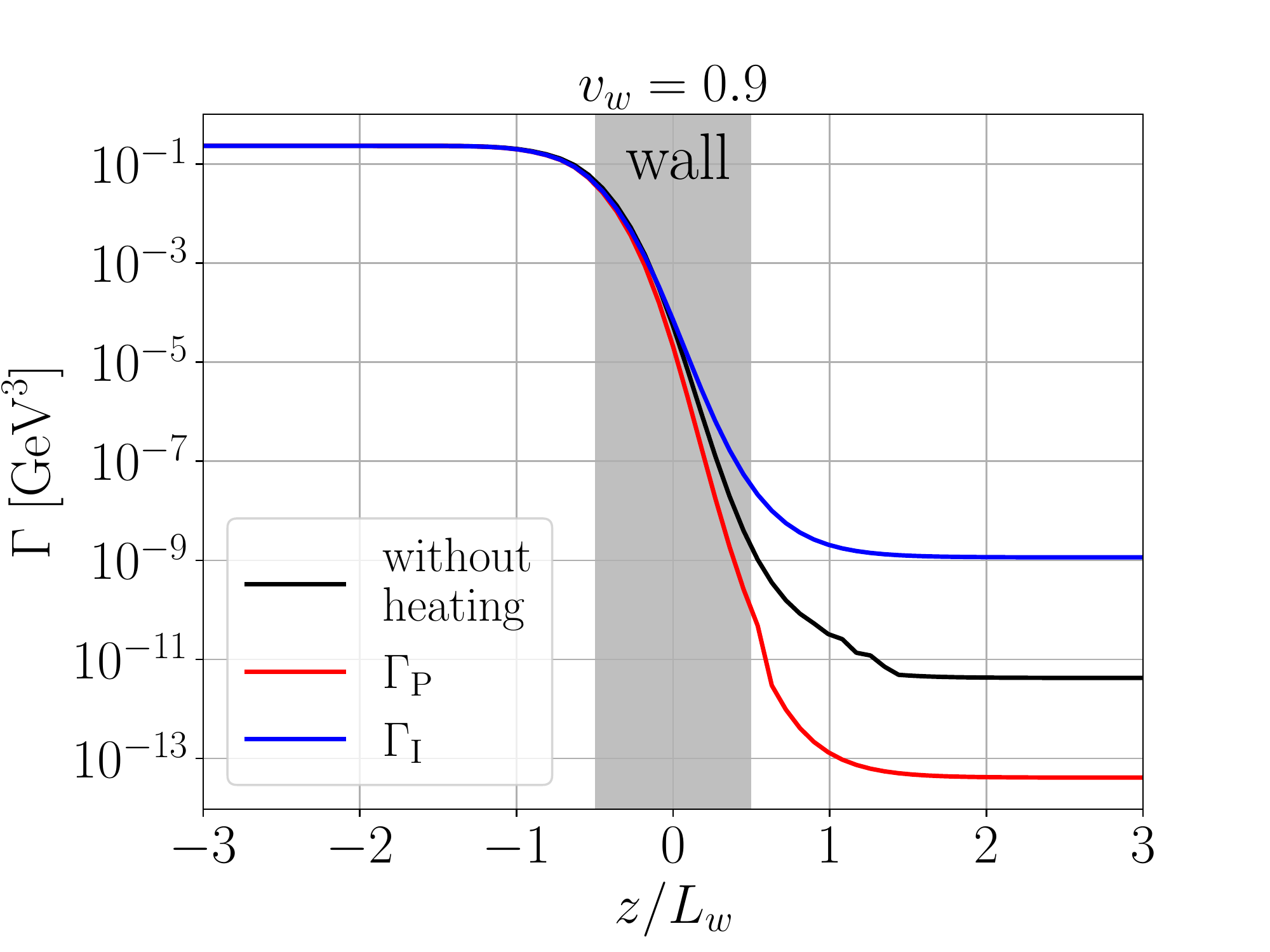}
	\end{minipage}}	
	\caption{Collision rates for the incident DM particle with $p_z^{\mathrm{ini}}=600~\mathrm{GeV}$. Red and blue lines are collision rates considering heating effects and black lines are collision rates without heating effects. Left panel is for $v_w=0.01$ and right panel is for $v_w=0.9$.}\label{Gamma}.
\end{figure}

We show in Fig.~\ref{Gamma} the collision rate $\Gamma_{\rm P(I)}(z,p_z)$ for $y_\chi=1$ and $p_z^{\mathrm{ini}}=600 ~\mathrm{GeV}$ where $p_z^{\mathrm{ini}}$ is the initial incident $z$ momentum towards the bubble wall. As we can see, the collision rates are different for each case due to the hydrodynamic effects. In the detonation case, the temperature in front of the bubble wall is the same, resulting in the same collision rate. However, heating effects behind the bubble wall cause $T_-$ to be higher than $T_+$. This induces a larger collision rate for $\Gamma_{\mathrm{I}}$. And because of $T_+=T_n<T_-$ and $\phi_->\phi_n$, we also have that $\Gamma_{\mathrm{P}}<\Gamma_n<\Gamma_{\mathrm{I}}$. For the deflagration of $v_w=0.01$, as the plasma in front of the bubble wall is heated, the collision rate is slightly enhanced. Due to that $T_-<T_n<T_+$ and $\phi_-<\phi_n$, one can get that $\Gamma_{\mathrm{I}}<\Gamma_{n}<\Gamma_{\mathrm{P}}$.  

For decay process, we define 
\begin{equation}
\begin{aligned}
g_\chi \int \frac{d p_x d p_y}{(2 \pi)^2} \mathbf{C}\left[f_\chi\right]&=-g_\chi \int \frac{d p_x d p_y}{(2 \pi)^2} d \Pi_{q^\mathcal{P}} d \Pi_{k^\mathcal{P}} \frac{(2 \pi)^4}{2 E_p^\mathcal{P}} \delta^{(4)}\left(p^\mathcal{P}+q^\mathcal{P}-k^\mathcal{P}\right)|\mathcal{M}|_{\chi \bar{\chi} \leftrightarrow \phi}^2 \\  & \quad \times \left[f_{\chi_p} f_{\bar{\chi}_q,+}^{\mathrm{eq}}-f_{\phi_k}^{\mathrm{eq}} \right] \,\,, \\
&=-g_\chi \int \frac{d p_x d p_y}{(2 \pi)^2} d \Pi_{q^\mathcal{P}} d \Pi_{k^\mathcal{P}} \frac{(2 \pi)^4}{2 E_p^\mathcal{P}} \delta^{(4)}\left(p^\mathcal{P}+q^\mathcal{P}-k^\mathcal{P}\right)|\mathcal{M}|_{\chi \bar{\chi} \leftrightarrow \phi}^2 \,\,,\\ & \quad \times \left[\mathcal{A}f_{\chi_p,+} f_{\bar{\chi}_q,+}^{\mathrm{eq}}-f_{\bar{\chi}_p}^{\mathrm{eq}} f_{\bar{\chi}_q}^{\mathrm{eq}} \right]\,\,,  \\
&\equiv \Gamma_{\rm P}(z,p_z) \mathcal{A}\left(z, p_z\right)-\Gamma_{\rm I}(z,p_z) \,\,,
\end{aligned}
\end{equation}
with
\begin{eqnarray}
\Gamma_{\rm I (P)}(z,E_p^{\mathcal{P}}) = -\int d p_x d p_y d q^{\mathcal P}	\frac{g_\chi g_{\bar\chi}m_\phi^2 q^{\mathcal P}}{ 128\pi^4E_p^\mathcal{P} E_q^{\mathcal P} p^\mathcal{P}  } f_{\bar{\chi}_{p(,+)}}^{\mathrm{eq}} f_{\bar{\chi}_{q(,+)}}^{\mathrm{eq}} \,\,,
\end{eqnarray}
where we have used for inverse decay $|\mathcal M|_{\chi \bar{\chi} \leftrightarrow \phi}^2=4y_\chi^2(p {\cdot} q)$ and then integrate over the phase space.

\subsection{Numerical results}
\subsubsection{Solutions to Boltzmann equations}
Due to energy and momentum conservation, along the paths on which both the transverse momentum $ p_{\perp}$ and the quantity $p_z^2+m_\chi^2(z)$ are constant, the differential operator simply reduces to a total derivative with respect to $z$ :
\begin{eqnarray}
\frac{p_z}{m_\chi} \frac{\partial}{\partial z}-\left(\frac{\partial m_\chi}{\partial z}\right) \frac{\partial}{\partial p_z} \quad \rightarrow \quad \frac{p_z}{m_\chi} \frac{d}{d z} \,\,,
\end{eqnarray}
which is called \emph{the method of characteristics}.

Then the Boltzmann equations can be transformed into the form of
\begin{eqnarray}
	\frac{d\mathcal{A} }{dz}=c(\mathcal{A},p_z,z)\,\,.
\end{eqnarray}
The explicit form of right-handed term can be extracted from Eqs.~\eqref{liou} and \eqref{coll}.

In order to solve for this differential equation we have to impose the boundary conditions. As the particles far in front of the bubble wall should be in thermal equilibrium, we can set $\mathcal{A}\left(z \ll-L_w, p_z>0\right)=1$. 
This uses the thermal-equilibrium condition that we discuss in Appendix B. 
We also assume an identical parallel wall at $z \gg L_w$. One can compute the solutions along the curves starting at $\left(z \ll-L_w, p_z>m_\chi^{\mathrm{ in}}\right)$ to find the values deep inside the bubble, at $\left(z \gg L_w, p_z>0\right)$. Then for particles which originate inside the bubble wall we set
\begin{eqnarray}
	\mathcal{A}\left(z \gg L_w, p_z\right)=\mathcal{A}\left(z \gg L_w,-p_z\right) \,\,.
\end{eqnarray}
This assumes that the flow form inside to outside comes from the other side of bubble wall.

\begin{figure}[htbp]
	\centering
	\subfigure{
		\begin{minipage}[t]{0.5\linewidth}
			\centering
			\includegraphics[scale=0.15]{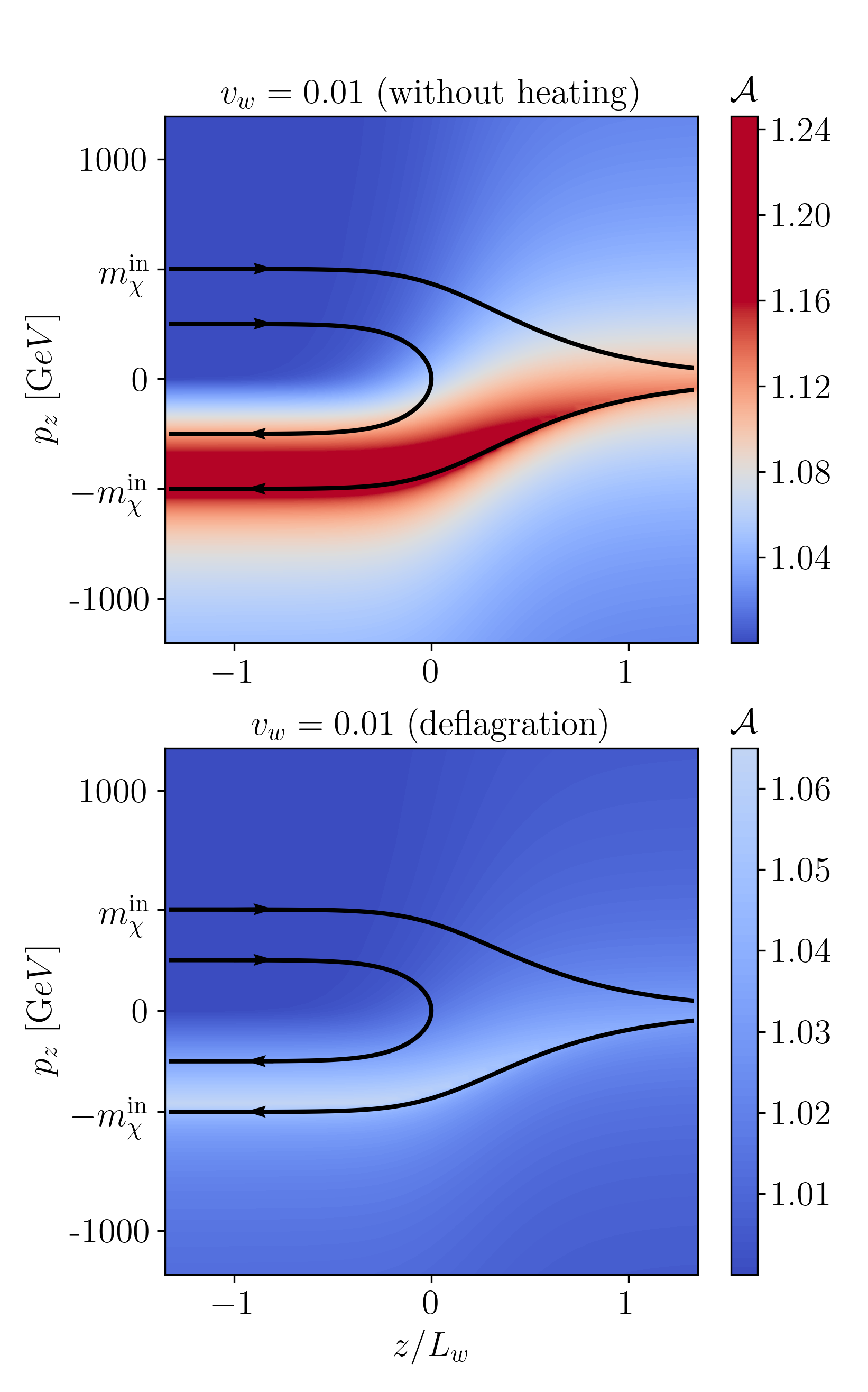}
	\end{minipage}}%
	\subfigure{
		\begin{minipage}[t]{0.5\linewidth}
			\centering
			\includegraphics[scale=0.15]{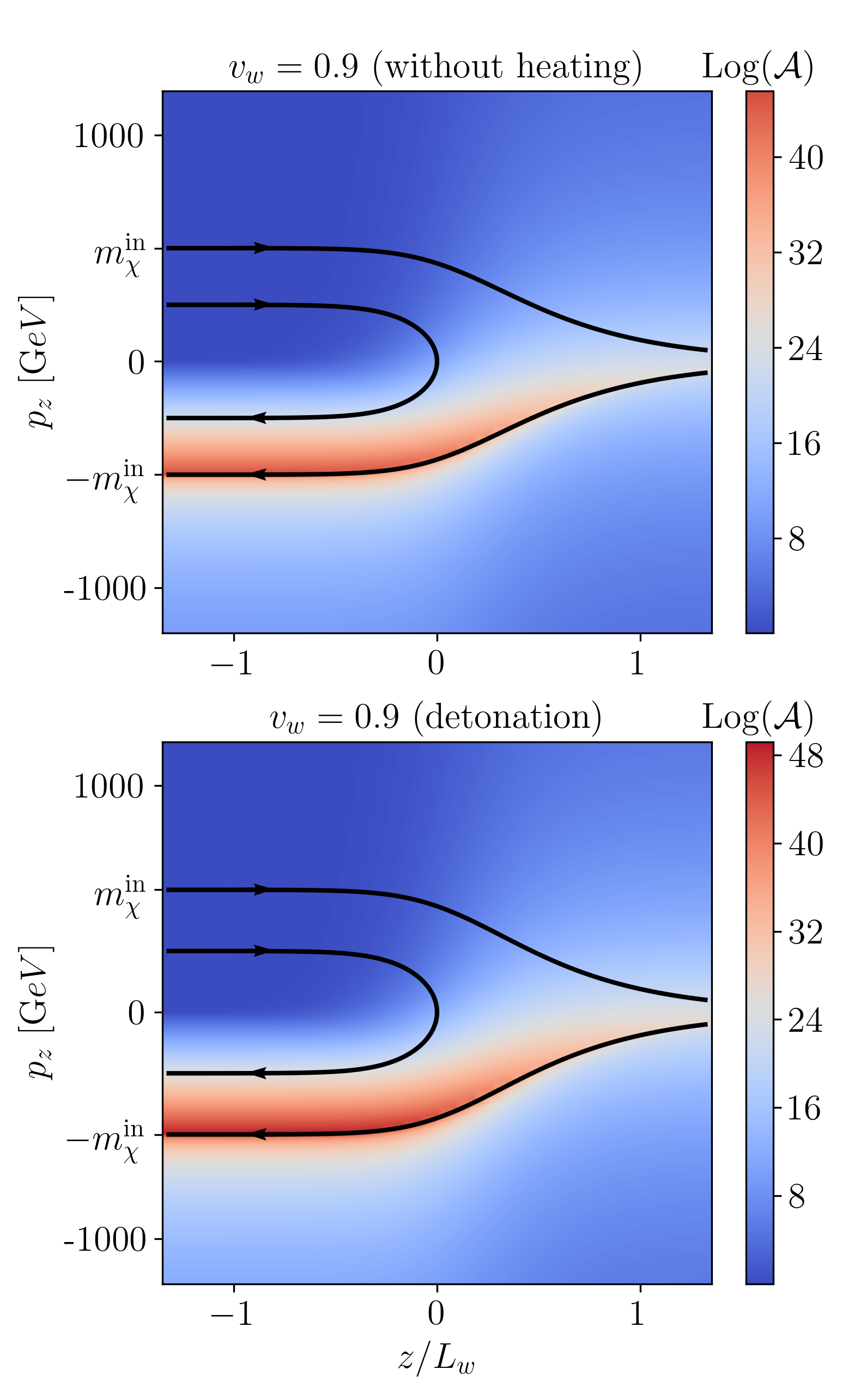}
	\end{minipage}}	
	\caption{Evolution of $\mathcal{A}(z,p_z)$ where we choose $y_\chi=1$. Left panels are for low velocity $v_w=0.01$ and right panels are for high velocity $v_w=0.9$.}\label{A}.
\end{figure}

After getting $\mathcal{A}(z,p_z)$, we can integrate and get the DM relic number density
\begin{equation}
\begin{aligned}
	n_{\chi}^{\rm in} =  
	\frac{T_+}{\gamma_w\tilde \gamma_+}\int_{0}^{\infty}\frac{dp_z}{(2\pi)^2}\mathcal A(z\gg L_w,p_z)\exp\left[\tilde \gamma_+\left(\tilde v_+ p_z-\sqrt{p_z^2+(m_{\chi}^{\mathrm in})^2}\right)/T_+\right]\left(\sqrt{p_z^2+(m_{\chi}^{\mathrm in})^2}+\frac{T_+}{\tilde \gamma_+}\right) \,\,,
\end{aligned}	 
\end{equation}
where the extra $1/\gamma_w$ accounts for the boost from the bubble wall frame into the bubble center frame.

We show the phase space enhancement factor $\mathcal A$ in Fig.~\ref{A}. The results are significantly different for cases with and without heating at $v_w=0.01$. Due to the heating effects in deflagration mode, the phase space enhancement factor is decreasing because of the smaller velocity in front of the bubble wall and smaller $m_\chi^{\mathrm{in}}$ behind the wall. Smaller $m_\chi^{\mathrm{in}}$ causes smaller perturbations for DM. For detonations the velocity is the same for both cases but $m_\chi^{\mathrm{in}}$ for detonations is slightly larger due to the heating effect, $\mathcal A$ is slightly enhanced.

One may worry about the too large $\mathcal A$ of our results. However, the $\mathcal A$ evaluates the non-equilibrium part of distribution function of DM. When the DM mass is much larger compared with the temperature, it almost fully decouples with the SM bath and we hope the non-equilibrium part should be large. This can be seen from the extreme case that the DM decouple with thermal bath immediately when the DM particles hit the bubble wall, in other words the DM is in a purely non-equilibrium state. Then we can ignore the collision term in the Boltzmann equation. Thus we have
\begin{eqnarray}
	\frac{d}{dz}(\mathcal A f^{\mathrm{eq}}_\chi)=0
\end{eqnarray}
along the flow path of DM. This gives us
\begin{eqnarray}
	\mathcal A\left(z \gg L_w, p_z\right) \exp \left(-\frac{\tilde \gamma_{+}(E-\tilde v_{+} p_z)}{T_{+}}\right) = 	\mathcal A\left(z \ll -L_w, p_z^{\mathrm{ini}}\right) \exp \left(-\frac{\tilde \gamma_{+}(E-\tilde v_{+} p_z^{\mathrm{ini}})}{T_{+}}\right)\,\,.
\end{eqnarray}
Since we have $\mathcal A\left(z \ll -L_w, p_z^{\mathrm{ini}}\right)=1$ and energy-conservation $p_z^2+m_\chi^2=\text{const.}$, we can get $(p_z^{\mathrm{ini}})^2=p_z^2+(m_\chi^{\mathrm{in}})^2$,  which gives us
\begin{eqnarray}
	\mathcal A\left(z \gg L_w, p_z\right) = \exp\left(\frac{\tilde \gamma_+ \tilde v_+(\sqrt{p_z^2+(m_\chi^{\mathrm{in}})^2}-p_z)}{T_+}\right)\,\,.
\end{eqnarray}
This can obviously be very large. Generally, the collisions with thermal bath help DM to return to thermal equilibrium, so we expect a smaller $\mathcal A$ when we include the collision terms.

\begin{figure}[htbp]
	\centering
	\subfigure{
		\begin{minipage}[t]{0.5\linewidth}
			\centering
			\includegraphics[scale=0.43]{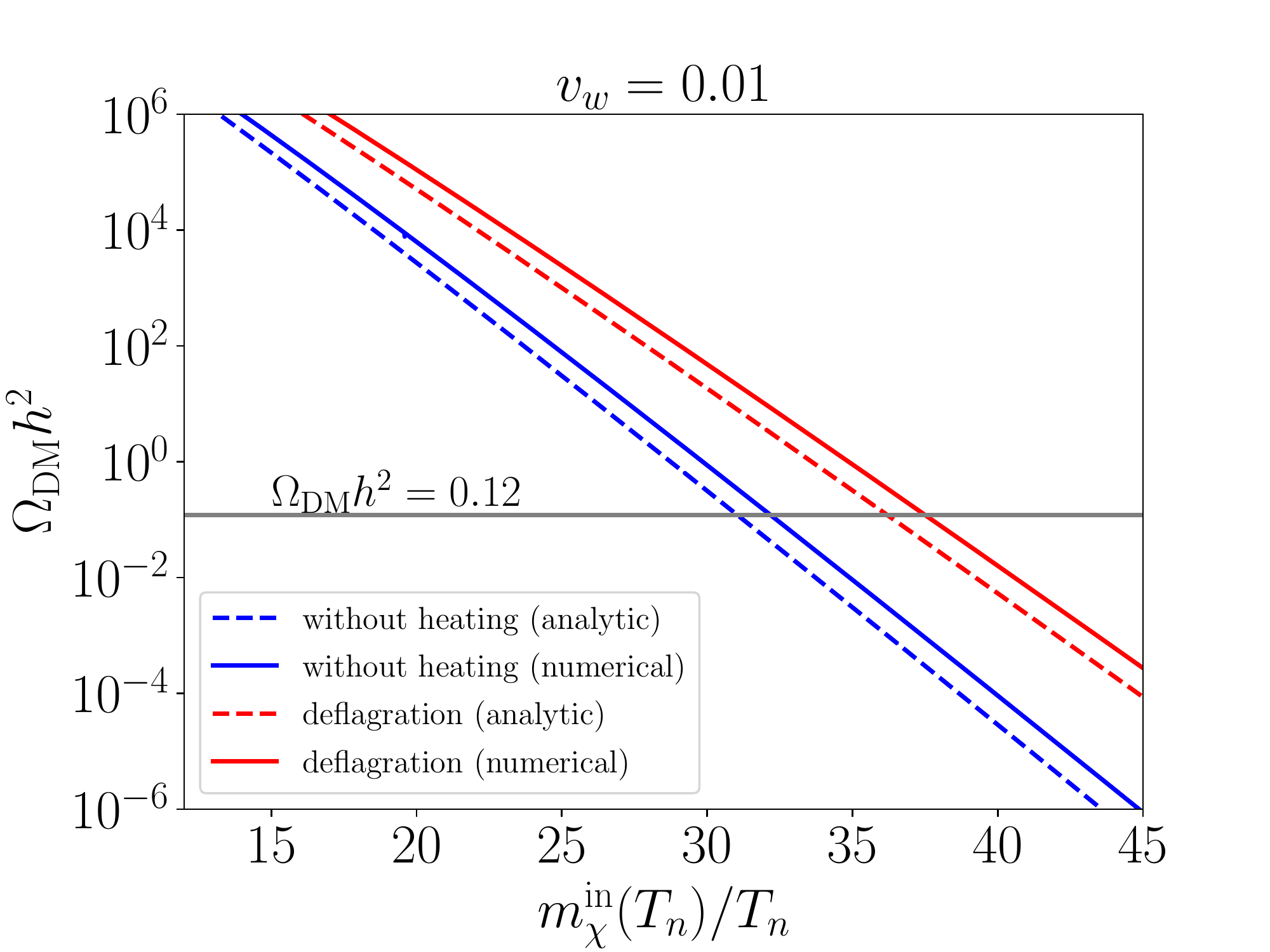}
	\end{minipage}}%
	\subfigure{
		\begin{minipage}[t]{0.5\linewidth}
			\centering
			\includegraphics[scale=0.43]{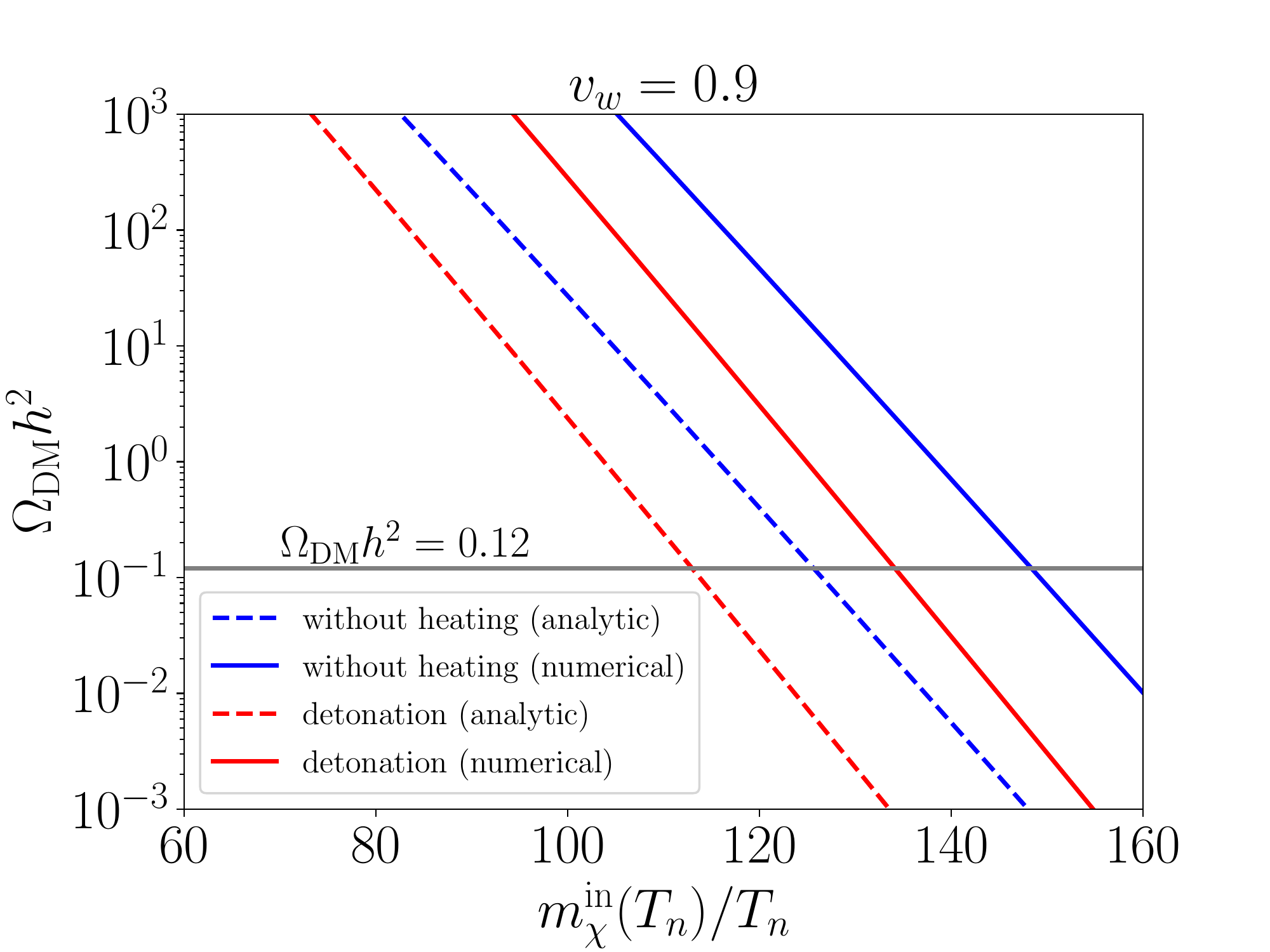}
	\end{minipage}}	
	\caption{$\Omega_{\mathrm{DM}}h^2$ as function of $m_\chi^{\mathrm{in}}(T_n)/T_n$ with $m_\chi^{\mathrm{in}}(T_n)\equiv y_\chi \phi_n$ with and without hydrodynamic heating effects. The analytic results are plotted by the dashed curves and numerical results are plotted by the solid curves. The blue curves represent cases without heating effects and the red curves represent the cases with the corresponding hydrodynamic mode.  }\label{oh2}.
\end{figure}

In the left panel of Fig.~\ref{oh2} we show the DM relic density for $BP_1$ where we choose $v_w=0.01$. Analytic results indicate that the DM relic density for deflagration mode is nearly 65 times greater than the one without heating, while numerical results suggest that the enhancement is almost 70 times. The calculations in analytic and numerical method differ by a factor of 3. For detonations in the right panel of Fig.~\ref{oh2} we have $\Omega_{\mathrm{DM}}^{(\mathrm{hy})} h^2/\Omega_{\mathrm{DM}}^{(0)} h^2=1/19$ in analytic method and $ \Omega_{\mathrm{DM}}^{(\mathrm{hy})} h^2/\Omega_{\mathrm{DM}}^{(0)} h^2=1/27$ in numerical method. The hydrodynamic effects of detonation are weaker than deflagration because of that for detonation the only difference is the temperature behind the bubble wall and so the DM field-dependent mass $m_\chi^{\mathrm{in}}$.

The analytic results of the deflagration case can be easily understood by the low-velocity approximation. We choose $y_\chi=2.8$ and thus $\Omega_{\mathrm{DM}}^{(0)}h^2=0.12$ for $BP_1$. For $BP_1$ we have $\alpha_n=0.23$ and $T_n=45.13~\mathrm{GeV}$. From Eq.~\eqref{lva}, we get
\begin{equation}
	\tilde v_-=0.01, \quad \tilde v_+ \simeq 0.003, \quad T_+ \simeq 45.13~\mathrm{GeV}, \quad T_-\simeq 33.67~\mathrm{GeV}
\end{equation}
then from Eqs.~\eqref{hyc}, \eqref{nana2} and \eqref{oh2hy} we can get
\begin{equation}
	\frac{\Omega_{\mathrm{DM}}^{(\mathrm{hy})}h^2}{\Omega_{\mathrm{DM}}^{(0)}h^2} = \frac{m_\chi^{\mathrm{in}}(T_-)n_\chi^{\mathrm{in}}(T_-)}{m_\chi^{\mathrm{in}}(T_n)n_\chi^{\mathrm{in}}(T_n)} \times \frac{T_n^3}{T_-^3} \simeq 67\,\,,
\end{equation}
which is consistent with our results. $n_\chi^{\mathrm{in}}(T_-)$ is the DM number density in the deflagration mode and $n_\chi^{\mathrm{in}}(T_n)$ is the number density without hydrodynamic effects. As we have mentioned below Eq.~\eqref{nana1}, the filtering effects mainly come from the competition between the kinetic energy $\tilde\gamma_+ T_+$ of the DM particle and the mass $m_\chi^{\mathrm{in}}$ deep inside the bubble. For the deflagration mode with $v_w=0.01$, the main influence comes from the variation of temperature across the bubble wall. In particular, for $BP_1$,
\begin{equation}
	\frac{m_\chi^{\mathrm{in}}(T_-)/T_+}{m_\chi^{\mathrm{in}}(T_n)/T_n} \simeq 0.88\,\,
\end{equation}
leads to
\begin{equation}
	n_\chi^{\mathrm{in}}(T_-)/n_\chi^{\mathrm{in}}(T_n) \simeq 31\,\,.
\end{equation}
We can see that the DM number density is enhanced due to the suppression of $m_\chi^{\mathrm{in}}/T$ which evaluates the competition between the DM kinetic energy and the DM mass deep inside the bubble. For the same reason, the relic density for detonation mode is suppressed. For detonation, only the temperature $T_-$ inside the bubble and the corresponding $m_\chi^{\mathrm{in}}(T_-)$ vary due to the hydrodynamic effects. Generally, the temperature behind the bubble wall rises in the case of detonation, $T_->T_n$. The DM mass behind the bubble wall is then enlarged. The kinetic energy does not vary, $\tilde \gamma_+ T_+ =\gamma_w T_n$. As a consequence, the DM relic abundance is suppressed. For the deflagration with $v_w \gtrsim 0.01$ and hybrid case, it is a little difficult to estimate the results from the analytic method. However, the calculation is still straightforward.

The numerical method of solving Boltzmann equations gives more accurate results than the analytic method. The reason is that the analytic method takes an assumption: the bubble wall has zero width and filtering process is instant.
This is not realistic in most phase transition processes. When DM particles penetrate into the bubbles, they experience interactions with other particles in the plasma. The temperature and velocity profile of SM plasma can influence the temperature and velocity of DM. This causes a redistribution of DM particles in the phase space. As a consequence, the DM number density is also influenced.

\subsubsection{Results for benchmark points}
We show the results for four benchmark points in Tabs.~\ref{01table} and \ref{09table} for $v_w=0.01$ and $v_w=0.9$, respectively. We fix the value of $m_\chi^{\mathrm{in}}(T_n)/T_n$ so that in the case without hydrodynamic effects we have $\Omega_{\mathrm{DM}}^{(0)} h^2=0.12$. When the value of $m_\chi^{\mathrm{in}}(T_n)/T_n=y_\chi \phi_n/T_n$ is fixed, the value of $y_\chi$ for a specific benchmark point is fixed, too. As we can see in Tab.~\ref{01table}, for $v_w=0.01$ the DM relic density is generally enhanced due to the heating effects. And the enhancement is stronger as the phase-transition strength $\alpha_n$ is larger. This is because that there is larger difference of temperature and velocity between the in front of and behind the bubble wall. For $BP_4$ with $v_w=0.01$, there is no deflagration mode for $\alpha_+>1/3$.

Tab.~\ref{09table} shows that, for detonation mode, the DM relic density is typically lower than that without heating, with differences of generally one or two orders of magnitude. The effects are slighter as the phase-transition is weaker. Notably, for $BP_4$ the case is actually a hybrid mode, resulting in extremely significant effects. This can be seen from Eq.~\eqref{Jouguet velocity} that for $BP_4$ we have $v_J(\alpha_n)\simeq 0.92$. Heating effects in hybrid mode are particularly large because both the front and back of the bubble wall experience significant changes. However, for the detonation mode the hydrodynamic effects work only behind the bubble wall and it influences only the field value or DM mass.  
\begin{table}[t]
	
	\centering
	
	\setlength{\tabcolsep}{3mm}
	
	\begin{tabular}{c|c|c|c|c}
		\hline\hline
		& \multicolumn{2}{c|}{analytic} & \multicolumn{2}{c}{numerical} \\
		\hline
		&  $m_\chi^{\mathrm{in}}(T_n)/T_n$ & $\Omega_{\mathrm{DM}}^{(\mathrm{hy})} h^2/\Omega_{\mathrm{DM}}^{(0)} h^2$ &$m_\chi^{\mathrm{in}}(T_n)/T_n$  &$\Omega_{\mathrm{DM}}^{(\mathrm{hy})} h^2/\Omega_{\mathrm{DM}}^{(0)} h^2$  \\
		\hline
		$BP_1$	&31 &66  &32 & 71 \\
		\hline
		$BP_2$	&  31.1 & 7.9 & 32.2 & 8.1\\
		\hline
		$BP_3$	& 30.8   & 778.8 & 31.9 & 858.5 \\
		\hline 
		$BP_4$	& $\star$  & $\star$  &$\star$  &$\star$  \\
		\hline\hline
	\end{tabular}
	\caption{Results for the four benchmark points for $v_w=0.01$. The $m_\chi^{\mathrm{in}}(T_n)/T_n$ is set by $\Omega_{\mathrm{DM}}^{(0)} h^2=0.12$. For $BP_4$ there is no deflagration mode because of the large phase-transition strength $\alpha_n$.}\label{01table}
\end{table}

\begin{table}[t]
	
	\centering
	
	\setlength{\tabcolsep}{3mm}
	
	\begin{tabular}{c|c|c|c|c}
		\hline\hline
		& \multicolumn{2}{c|}{analytic} & \multicolumn{2}{c}{numerical} \\
		\hline
		&  $m_\chi^{\mathrm{in}}(T_n)/T_n$ & $\Omega_{\mathrm{DM}}^{(\mathrm{hy})} h^2/\Omega_{\mathrm{DM}}^{(0)} h^2$ &$m_\chi^{\mathrm{in}}(T_n)/T_n$  &$\Omega_{\mathrm{DM}}^{(\mathrm{hy})} h^2/\Omega_{\mathrm{DM}}^{(0)} h^2$  \\
		\hline
		$BP_1$	&125.3 &1/19  &147.8 & 1/27 \\
		\hline
		$BP_2$	&  125.9 & 1/7 & 148.7 & 1/9\\
		\hline
		$BP_3$	& 124.6   & 1/10 & 147.3 & 1/12 \\
		\hline 
		$BP_4$	&  123.8 & $1/(1.2\times 10^{13})$   & 146.5  & $1/(2.2\times 10^{15})$ \\
		\hline\hline
	\end{tabular}
	\caption{Results for the four benchmark points for $v_w=0.9$. $BP_{1,2,3}$ are detonation cases and $BP_4$ is the hybrid case. }\label{09table}
\end{table}

\section{phase transition gravitational wave signals of the filtered dark matter}\label{gw}
\begin{figure}[htbp]
	\begin{center}
		\includegraphics[scale=0.8]{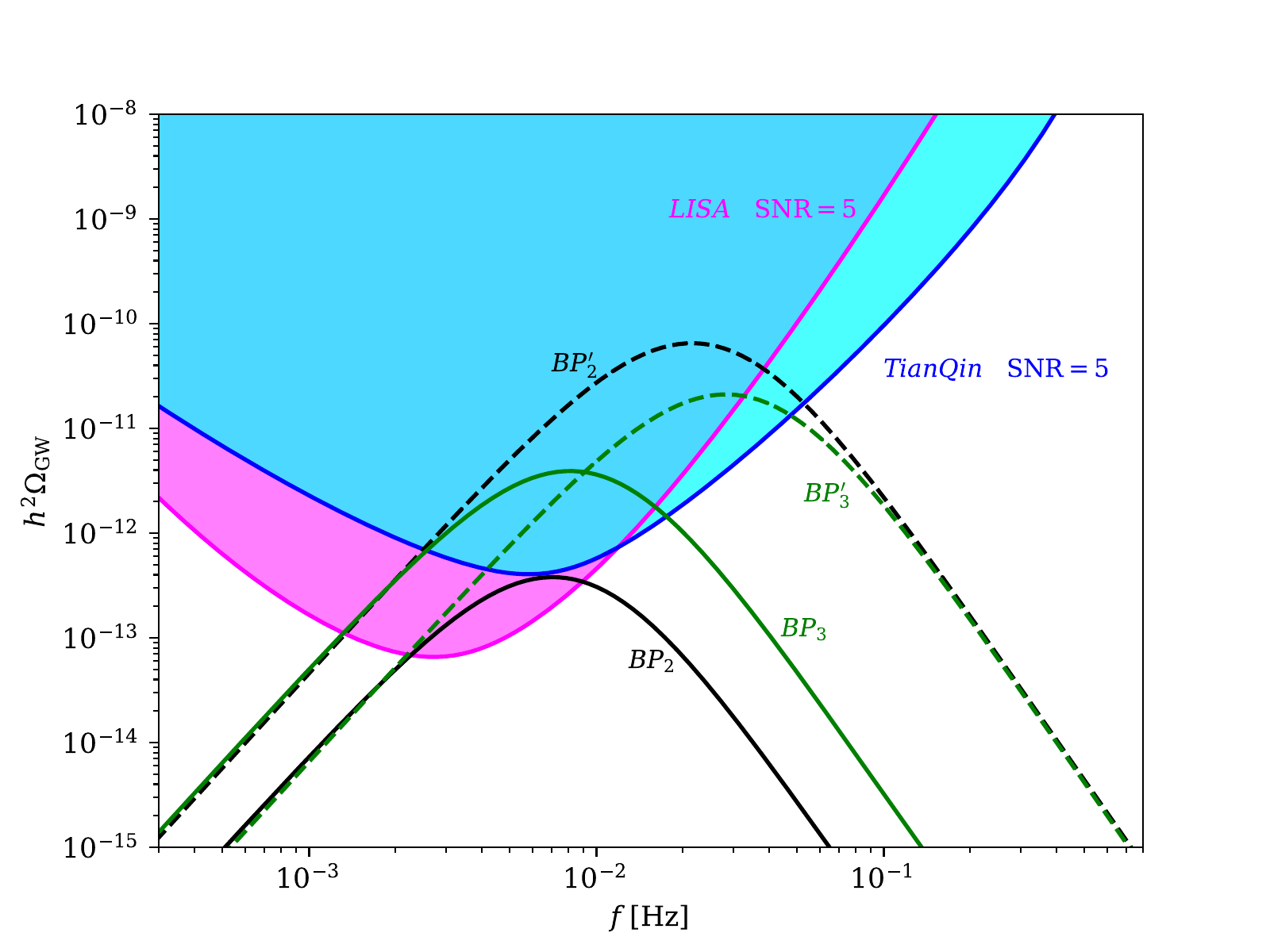}
		\caption{Phase transition GW spectra for $BP_{2,3}$ and $BP_{2,3}'$. The colored regions represent the sensitivity curves for LISA and TianQin with SNR= 5
			for $10^8$ s observation time, respectively.}
		\label{fig_gwfdm}
	\end{center}
\end{figure}

Naturally, this filtered DM mechanism could be detected by phase transition GW signals from bubble collision~\cite{Witten:1984rs, Hogan:1986qda, Kamionkowski:1993fg},  turbulence~\cite{Kamionkowski:1993fg} and sound wave~\cite{Hindmarsh:2017gnf} during a SFOPT.
We just show the phase transition GW spectra for $BP_2$ (solid black line) and $BP_3$ (solid green line) with $v_w=0.9$ in Fig.~\ref{fig_gwfdm} for simplicity.
The colored regions represent the sensitivity curves for LISA~\cite{LISA:2017pwj} and TianQin~\cite{TianQin:2015yph} with the signal-to-noise ratio (SNR) about 5
for $10^8$ s observation time, respectively.  The GW spectra are sensitive to bubble wall velocity. For example, for $v_w=0.01$, the GW spectrum is much weaker. The corresponding SNR will be lower than 5.
From Fig.~\ref{fig_gwfdm}, we could see that the future GW detectors LISA, TianQin~\cite{Liang:2022ufy} 
could detect this new DM mechanism with SNR larger than 5.   Taiji~\cite{Hu:2017mde}, BBO~\cite{Corbin:2005ny},
DECIGO~\cite{Seto:2001qf}, Ultimate-DECIGO~\cite{Kudoh:2005as} could  also detect this new DM mechanism by GW signals.
\begin{table}[t]
	\centering
	
	\setlength{\tabcolsep}{3mm}
	
	\begin{tabular}{c|c|c|c|c|c|c|c}
		\hline\hline
		& $D$ & $C$ & $\lambda$ &$T_n $~[GeV] &$\alpha_n$ & $\beta/H_n$ &$\phi_n/T_n$\\
		\hline
		$BP_2'$	&2.1 &0.036  &0.01 & 17.34 & 1.25 &12226  &10 \\
		\hline
		$BP_3'$	&5.8 &0.04  &0.01 & 10.3 & 4.48 &27037  & 11.1\\
		\hline\hline
	\end{tabular}
	\caption{The modified benchmark parameters that fulfill $\Omega_{\mathrm{DM}}^{(0)}h^2=0.12$ based on $BP_2$ and $BP_3$.}\label{ptablep}
\end{table}

It may be difficult to discuss the impact of hydrodynamic effects of the filtered DM mechanism on the GW spectra because there are four free parameters in our model: $D$, $C$, $\lambda$, and the Yukawa coupling $y_\chi$. However, as an example, we can fix three of them and take the parameter $D$, which directly varies $\phi_n$ and $T_n$ as the free parameter. The $y_\chi$ is chosen so that $m_\chi^{\mathrm{in}}(T_n)/T_n=y_\chi \phi_n/T_n$ which fulfills $\Omega_{\mathrm{DM}}^{(\mathrm{hy})}h^2=0.12$ for $BP_i$~($i$=2, 3). We introduce new benchmark points $BP_i'$ to compare with $BP_i$. The other three parameters except $D$ are the same as $BP_i$. We choose the detonation case $v_w=0.9$ for simplicity. As we have mentioned in Sec~\ref{sa}, for detonation the hydrodynamic heating effects only influence the temperature behind the bubble wall, then we get $m_\chi^{\mathrm{in}}(T_-) >m_\chi^{\mathrm{in}}(T_n) $. In order to fulfill the relic density for $BP_i'$, which does not consider the heating effects, we have to impose the condition:
\begin{equation}
	\Omega_{\mathrm{DM}}^{(0)}h^2=0.12\,\,.
\end{equation}
We scan the parameter region of $D$ for $BP_i'$ and find the value of $D$ to satisfy the above condition. Other phase-transition parameters are shown in Tab.~\ref{ptablep}. The GW spectra of $BP_2'$ and $BP_3'$ are shown in terms of the dashed lines in the Fig.~\ref{fig_gwfdm}. We can see that the GW spectra for filtered DM with and without hydrodynamic effects are dramatically different with those for filtered DM without hydrodynamic effects. More detailed discussions on the GW detection
are left in our future work.

\section{Conclusion}\label{conclusion}

We have found that the hydrodynamic effects play essential roles in the filtered DM mechanism for the reason that the DM relic density is sensitive to the hydrodynamic modes, the bubble wall velocity, the temperature, and velocity profile in the vicinity of bubble wall. 
On one hand, to produce efficient filtered effects, the parameter $\phi/T$ is usually very large, which generally corresponds to large phase transition strength $\alpha$. And for large $\alpha$, the hydrodynamic effects are significant since in this case, the temperature and velocity differences in front of and behind the bubble wall are significant.
On the other hand, the bubble wall velocity and the hydrodynamic modes are also essential to the final relic DM density. 
For the deflagration mode with low  bubble wall velocity, the hydrodynamic effects significantly enhance the relic density. In contrast, for the
detonation mode, the relic density is obviously reduced. 
For the hybrid mode, the hydrodynamic correction is extremely large.
Our study could be applied in various concrete filtered DM models and general phase transition process in the early Universe.

%

\begin{acknowledgments}
The authors thank Chang Sub Shin for helpful correspondence. We acknowledge Zheng-Cheng Liang to provide the updated sensitivity curve. This work is supported by the National Natural Science Foundation of China (NNSFC) under Grant No. 12205387. 
\end{acknowledgments}

\appendix
\section{PHASE TRANSITION PARAMETERS}\label{PTa}
Here we present the results from Ref.~\cite{Ellis:2020awk} which describe the semi-analytical calculation of phase transition parameters in the toy model,
\begin{eqnarray}
	\alpha = \frac{15(\sqrt{9-4\delta}+3)(-4\delta+3\sqrt{9-4\delta}+9)C^2(D\lambda-\delta C^2)}{2^3\pi^2\sqrt{9-4\delta}g_\star \lambda^3}\,\,,
\end{eqnarray}
and
\begin{eqnarray}
	\frac{\beta}{H} = \frac{192\pi\sqrt{\delta}[-\beta_1(\delta+6)+\beta_2\delta(\delta-10)+\beta_3\delta^2(3\delta-14)](D\lambda-\delta C^2)}{243(\delta-2)^3C\lambda^{3/2}} 
\end{eqnarray}
with $\beta_1$, $\beta_2$, and $\beta_3$ given in Eq.~\eqref{Sana} and $\delta$ defined in Eq.~\eqref{delta}.

\begin{figure}[htbp]
	\begin{center}
		\includegraphics[width=0.7\linewidth]{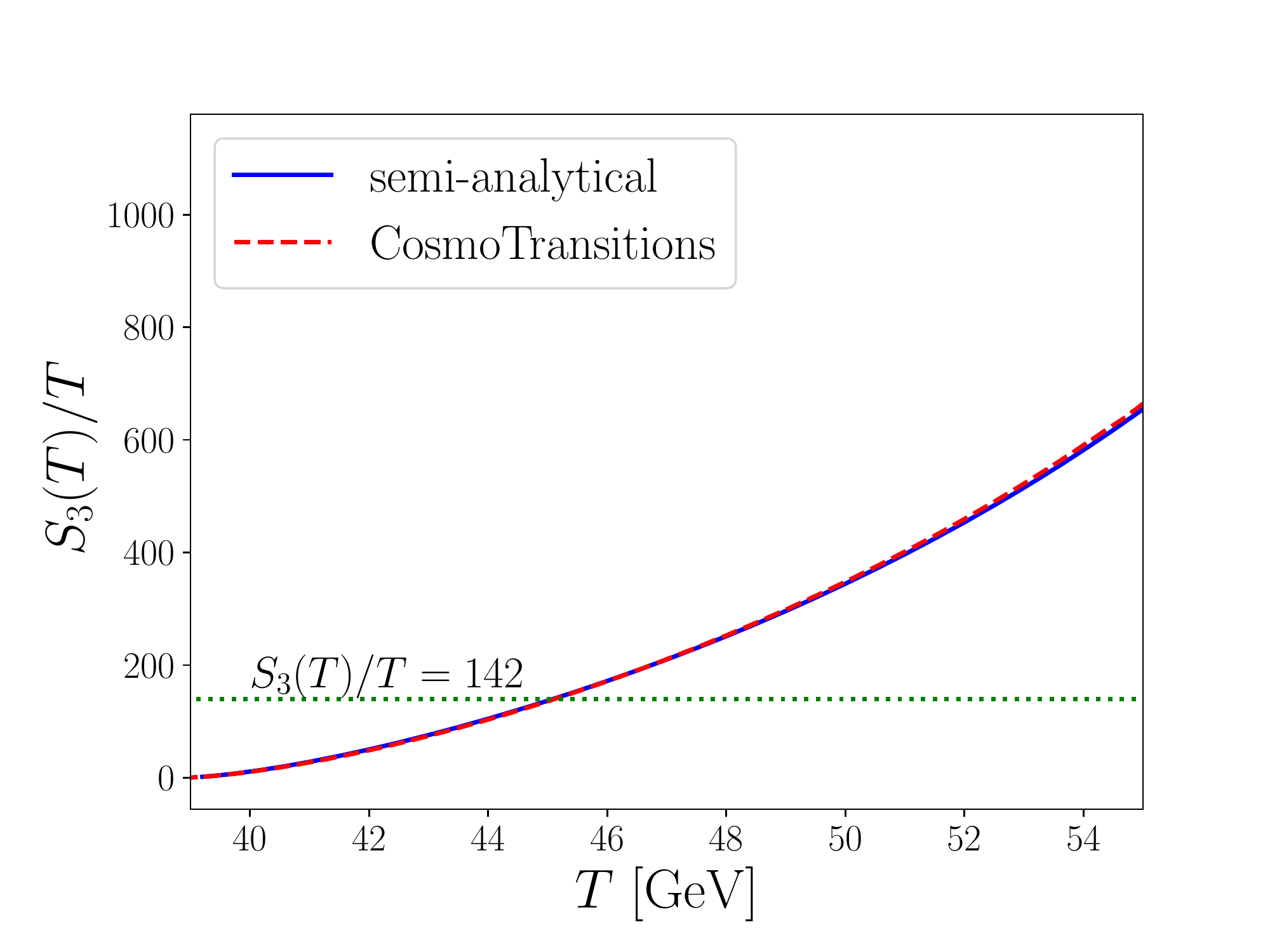}
		\caption{$S_3(T)/T$ as function of temperature $T$ for $BP_1$ in semi-analytical method and in Cosmotransitions.}
		\label{S3T}
	\end{center}
\end{figure}

In Fig.~\ref{S3T} we show the bounce action $S_3(T)/T$ as function of temperature. We can see that the the CosmoTransitions gives the same results as the semi-analytical calculation.

\section{THERMAL EQUILIBRIUM}\label{kd}
In the previous calculation, we have assumed that the DM has a thermal equilibrium distribution in front of the bubble wall. This requires that the reflected DM should return to the same momentum as $p \sim T$ of the thermal plasma. In order to evaluate this, we have to evaluate the momentum exchange between reflected DM and the DM that is in thermal equilibrium~\cite{Kawana:2022lba,Baldes:2022oev}.

In the false vacuum, the DM is massless. Some DM particles are reflected by the bubble wall and will be out of equilibrium.
By denoting the momentum of a particle in the plasma frame as $p^{\mathcal{P}}=\left(E^\mathcal{P}, p_x^\mathcal{P}, p_y^\mathcal{P}, p_z^\mathcal{P}\right)$ with $E^\mathcal{P}=$ $\sqrt{m_\chi^2+(p_x^{\mathcal{P}})^2+(p_y^{\mathcal{P}})^2+(p_z^{\mathcal{P}})^2}$, the transformed momentum in the wall frame is
\begin{eqnarray}
p^{}=\left(E^{}, p_x^{}, p_y^{}, p_z^{}\right)=\left(\frac{E^\mathcal{P}+v_w p_z^\mathcal{P}}{\sqrt{1-v_w^2}}, p_x^\mathcal{P}, p_y^\mathcal{P}, \frac{p_z^\mathcal{P}+v_w E^\mathcal{P}}{\sqrt{1-v_w^2}}\right)\,\,.
\end{eqnarray}
After DM reflects off the wall we have $p_z^{} \rightarrow$ $\tilde{p}_z^{}=-p_z^{}$, then the reflected momentum of DM in the plasma frame becomes
\begin{eqnarray}
\tilde{p}^\mathcal{P}=\left(\tilde{E}^\mathcal{P}, \tilde{p}_x^\mathcal{P}, \tilde{p}_y^\mathcal{P}, \tilde{p}_z^\mathcal{P}\right)=\left(\frac{E^{}+v_w p_z^{}}{\sqrt{1-v_w^2}}, p_x^{}, p_y^{}, \frac{-p_z^{}-v_w E^{}}{\sqrt{1-v_w^2}}\right)\,\,.
\end{eqnarray}

Then the $\tilde{E}^\mathcal{P}$ is given by
\begin{eqnarray}
\tilde{E}^\mathcal{P}=\frac{\left(1+v_w^2\right) E^\mathcal{P}+2 v_w p_z^\mathcal{P}}{1-v_w^2}\,\,,
\end{eqnarray}
and $\tilde p_z^\mathcal{P}$ by
\begin{eqnarray}
	\tilde p_z^\mathcal{P} = -\frac{\left(1+v_w^2\right) p_z^\mathcal{P}+2 v_w E^\mathcal{P}}{1-v_w^2}\,\,,
\end{eqnarray}
from which the momentum exchange in a single collision of DM in the plasma frame is 
\begin{eqnarray}
\delta p_z^\mathcal{P}=\tilde{p}_z^\mathcal{P}-p_z^\mathcal{P}=-2 v_w \frac{v_w p_z^\mathcal{P}+E^\mathcal{P}}{1-v_w^2}\,\,.
\end{eqnarray}

The thermal plasma has $|\mathbf{p}^\mathcal{P}| \sim \mathcal{O}(T)$. The reflected $\chi$ and the $\chi$ that is in thermal equilibrium have momenta
\begin{eqnarray}
p_1^\mathcal{P}=(T+\frac{2 v_w T}{(1-v_w)T}, 0,0,-T-\frac{2 v_w T}{(1-v_w)T}), \quad p_2^\mathcal{P}=(T, 0,0, T)\,\,,
\end{eqnarray}
respectively. The momentum of the c.m. frame is then
\begin{eqnarray}
p_{\mathrm{c.m.}}^\mathcal{P}=p_1^\mathcal{P}+p_2^\mathcal{P}=\left(\frac{2}{1-v_w} T, 0,0, \frac{-2 v_w}{1-v_w} T\right)\,\,.
\end{eqnarray}

Hence, the velocity of the c.m. frame relative to the plasma frame is $v_{\mathrm{c.m.}}=-v_w$ and the Mandelstam variable $\hat{s}$ of the scattering is
\begin{eqnarray}
\hat{s}=4 T^2 \frac{1+v_w}{1-v_w} \,\,.
\end{eqnarray}
The elastic collision in the c.m. frame is represented as $p_1^{\prime}+p_2^{\prime} \rightarrow p_3^{\prime}+p_4^{\prime}$ with
\begin{eqnarray}
\begin{aligned}
	& p_1^{\prime }=\left(E_{\mathrm{c.m.}}, 0,0,-E_{\mathrm{c.m.}}\right), \quad p_2^{\prime }=\left(E_{\mathrm{c.m.}}, 0,0, E_{\mathrm{c.m.}}\right), \\
	& p_3^{\prime }=\left(E_{\mathrm{c.m.}}, 0,-E_{\mathrm{c.m.}} \sin \theta,-E_{\mathrm{c.m.}} \cos \theta\right), \quad p_4^{\prime }=\left(E_{\mathrm{c.m.}}, 0, E_{\mathrm{c.m.}} \sin \theta, E_{\mathrm{c.m.}} \cos \theta\right)
\end{aligned}
\end{eqnarray}
with $E_{\mathrm{c.m.}}=p_{\mathrm{c.m.}}=\sqrt{\hat s}/2$.

The momentum loss in such a collision in the plasma frame is
\begin{eqnarray}
	\delta p_{\chi}=p^\mathcal{P}_{1z}-p^\mathcal{P}_{3z}=\gamma_{\mathrm{c.m.}} \left(p_{1 z}^{\prime}-p_{3 z}^{\prime}\right) =-\frac{\gamma_{\mathrm{c.m.}}\hat t}{2E_{\mathrm{c.m.}}}\,\,,
\end{eqnarray}
where we have Lorentz transformed between the c.m. and plasma frames and we use $\hat t =(p_1^{\prime }-p_3^{\prime })^2= -2E_{\mathrm{c.m.}}^2(1-\mathrm{cos}\theta)$.
Now we can derive the momentum-loss rate
\begin{eqnarray}
	\begin{aligned}
		\frac{d \log \left(p_{\chi}\right)}{d t} & \approx \frac{n_{\chi} v_{\text{Møl}}}{p_{\chi}} \int_{-4 E_{\mathrm{c.m.}}^2}^0 d \hat{t} \frac{d \sigma}{d \hat{t}} \delta p_{\chi} \,\,,\\
		& \approx-\frac{\gamma_{\mathrm{c.m.}}n_{\chi} v_{\text{Møl}}}{2 E_{\mathrm{c.m.}} T} \int_{-4 E_{\mathrm{c.m.}}^2}^0 d \hat{t} \frac{d \sigma}{d \hat{t}} \hat{t}\,\,,
	\end{aligned}
\end{eqnarray}
where $v_{\text{Møl}} \simeq 2$ is the Møller velocity and
\begin{eqnarray}
n_\chi=g_\chi \frac{3 \zeta_3}{4 \pi^2} T^3, \quad \frac{d \sigma}{d \hat{t}}=\frac{|i \mathcal{M}|^2}{16 \pi \hat{s}^2}=\frac{y_\chi^4}{16 \pi \hat{s}^2}\,\,.
\end{eqnarray}
Then the thermal equilibrium condition will be
\begin{eqnarray}
	\frac{d \log \left(p_{\chi}\right)}{d t}>H \rightarrow y_\chi \gtrsim 5.4\times 10^{-4} \left(\frac{g_\star}{120}\right)^{1/8} \left(\frac{2}{g_\chi}\right)^{1/4}\left(\frac{T_n}{100~\mathrm{GeV}}\right)^{1/4}(1+v_w)^{1/4}\,\,.
\end{eqnarray}

Besides this, we also require that the DM reflected off should be quickly annihilate away, then we have~\cite{Baker:2019ndr}
\begin{eqnarray}
	y_\chi \gtrsim\left(8 \times 10^{-4}\right)\left(\frac{g_*}{120}\right)^{1 / 8}\left(\frac{2}{g_\chi}\right)^{1/4}\left(\frac{T_n}{1~ \mathrm{TeV}}\right)^{1 / 4}\left(\frac{\log 36 T_n^2 / m_\phi^2}{\log 36}\right)^{-1 / 4}\,\,.
\end{eqnarray}

	

\bibliographystyle{apsrev}
\bibliography{filterdm}
\end{document}